\documentclass[12pt,preprint]{aastex}

\def\p{\partial}

\def\nab{\mbox{\boldmath $\nabla$}}

\def\rb{\bar{\rho}}
\def\tb{\bar{T}}
\def\sb{\bar{S}}

\def\vph{\hat{v}_{\phi}}

\def\vrr{\tilde{v}_r} 
\def\vtr{\tilde{v}_{\theta}}
\def\vphr{\tilde{v}_{\phi}}
\def\vvr{\tilde{v}}
\def\brr{\tilde{B}_r} 
\def\btr{\tilde{B}_{\theta}}
\def\bphr{\tilde{B}_{\phi}}
\def\bbr{\tilde{B}}

\newcommand{\cross}{\mbox{\boldmath $\times$}}
\newcommand{\BB}{{\bf B}}
\newcommand{\uvr}{\mbox{\boldmath ${\bf e}_r$}}
\newcommand{\uvt}{\mbox{\boldmath ${\bf e}_\theta$}}
\newcommand{\uvp}{\mbox{\boldmath ${\bf e}_\phi$}}

\begin{document}

\title{Simulations of core convection in rotating A-type stars:
Magnetic dynamo action}

\author{Allan Sacha Brun$^{1,2}$, Matthew K. Browning$^2$ and Juri Toomre$^2$}

\affil{$^1$DSM/DAPNIA/SAp, CEA Saclay, 91191 Gif sur Yvette, France, \\
$^2$JILA and Department of Astrophysical and Planetary Sciences,
University of Colorado, Boulder, CO 80309-0440}

\begin{abstract}

Core convection and dynamo activity deep within rotating A-type stars of 2
solar masses are studied with 3--D nonlinear simulations. Our modeling
considers the inner 30\% by radius of such stars, thus capturing within a
spherical domain the convective core and a modest portion of the
surrounding radiative envelope.  The magnetohydrodynamic (MHD) equations
are solved using the anelastic spherical harmonic (ASH) code to examine
turbulent flows and magnetic fields, both of which exhibit intricate time
dependence.  By introducing small seed magnetic fields into our progenitor
hydrodynamic models rotating at one and four times the solar rate, we
assess here how the vigorous convection can amplify those fields and
sustain them against ohmic decay.  Dynamo action is indeed realized,
ultimately yielding magnetic fields that possess energy densities
comparable to that of the flows.  Such magnetism reduces the differential
rotation obtained in the progenitors, partly by Maxwell stresses that
transport angular momentum poleward and oppose the Reynolds stresses in the
latitudinal balance. In contrast, in the radial direction we find that the
Maxwell and Reynolds stresses may act together to transport angular
momentum. The central columns of slow rotation established in the
progenitors are weakened, with the differential rotation waxing and waning
in strength as the simulations evolve.  We assess the morphology of the
flows and magnetic fields, their complex temporal variations, and the
manner in which dynamo action is sustained.  Differential rotation and
helical convection are both found to play roles in giving rise to the
magnetic fields.  The magnetism is dominated by strong fluctuating fields
throughout the core, with the axisymmetric (mean) fields there relatively
weak.  The fluctuating magnetic fields decrease rapidly with radius in the
region of overshooting, and the mean toroidal fields less so due to
stretching by rotational shear.

\end{abstract}

\section{INTRODUCTION}

\subsection{Surface Magnetism of A-type Stars}

The magnetic Ap stars have been objects of intense scrutiny for much of
the past century.  These stars are broadly characterized by strong spectral
lines of some elements (mainly Si and some rare earths), variability on
timescales of days to decades, and surface magnetic fields as strong as
tens of kG (see Wolff 1980 for a review).  Extensive observations, ranging
from the first analyses of the Ap star $\alpha^2$CVn by Maury (1897) to
recent surveys by Hubrig, North \& Mathys (2000), have painted a fairly
detailed picture of the many surface pathologies exhibited by these stars,
and have provided important clues about how the abundance features and
surface magnetism may arise.  Yet major puzzles remain.  We begin here by
outlining the major observational features of such stars, which serve to
motivate and guide the work described here.

Observations of the Zeeman effect in magnetic Ap stars suggest that the
surface fields are variable in apparent strength, that most exhibit
periodic reversals in polarity along the line of sight, and that they are
of large spatial scale.  The commonly accepted framework for the
interpretation of these observations is the ``rigid rotator'' model, in
which a global scale field is taken to have an axis of symmetry inclined at
some angle with respect to the rotation axis (e.g., Stibbs 1950; Mestel \&
Moss 1977; Moss, Mestel, \& Tayler 1990; Mestel 1999).  In this model,
variations in the apparent field strength are simply a consequence of the
star's rotation, as the magnetic axis continually changes its orientation
with respect to the line of sight.  Likewise, variations in elemental
abundance measurements are thought to result from viewing large patches of
those elements as they rotate in and out of view.

The geometry of the surface magnetic field has been interpreted as being
predominantly dipolar.  Quadrupolar and higher-order field components can
have only a small influence on integrated Zeeman measurements of the
longitudinal (line of sight) field component: Thus if the field was mainly
quadrupolar, the total field would have to be quite high (of order 20-40
kG) to yield commonly observed values for the longitudinal field component
(1-2 kG).  Such large values of the total field are ruled out for most Ap
stars by measurements or non-detections of resolved Zeeman splitting, which
is sensitive to the total surface field independent of direction.  However,
a field that is purely dipolar cannot account for the exact patterns of
variation observed for the longitudinal and total field, suggesting that
the surface magnetism does have some higher-order component (e.g., Borra
1980).  Very recently, extensive high-resolution spectropolarimetric
observations have begun to yield more direct constraints on the geometry of
the surface magnetic fields. Kochukhov et al. (2004) infer from line
profiles in all four Stokes parameters that the surface magnetic field of
53 Cam is quite complex in structure, with high-order multipoles ($l=10$
and greater) contributing strongly to the total field.  

A few broad characterizations of the extensive observations of magnetic Ap
star properties can be made.  The most striking, as noted by many authors
(e.g., Mestel 1999; Borra \& Landstreet 1980) is the gross
anti-correlation between rotation rate and magnetic field strength:
the magnetic Ap stars are preferentially much slower rotators than A-type
stars with no observed field.  Some exhibit variations with periods of
decades, which may imply very slow rotation rates indeed. However, there
are some magnetic Ap stars that have rotational velocities well in excess
of 100 km s$^{-1}$, so slow rotation does not appear to be an absolute
prerequisite for surface magnetism.  Within the class of magnetic Ap stars,
Hubrig, North \& Mathys (2000) see some evidence for a weak
correlation between rotation and magnetic flux, in that
shorter-period Ap stars exhibit marginally stronger fields. They also found
that only Ap stars that have completed at least one third of their
main-sequence lifetimes show magnetism, though younger magnetic Ap stars
have been observed by other authors (e.g., Bagnulo et al. 2003, who find
that HD 66318 has completed only about 16\% of its main-sequence liftime).
Finally, one of the most puzzling observational facts concerning these
stars is their relative rarity: only about 10\% of stars of the appropriate
spectral type are observably magnetic (e.g., Moss 2001).  There appears to
be no set of stellar parameters that is a sufficient condition for the
presence of magnetism in any given A-type star.

\subsection{Possible Origins of the Magnetism}

The central question raised by the extensive observational data
is most simply: What is the origin of the magnetism?  Two major theories have
emerged that seek to account for the observed fields.

The ``fossil'' theory suggests that the fields are relics of
the primordial field that threaded the interstellar gas out of which the
stars formed.  Ohmic decay times in the stable radiative envelopes of
A-stars are very long, so the primordial field, sufficiently concentrated
by the star formation process, might well survive through most or all of
such stars' main-sequence lifetimes.  In the fossil theory, the slow
rotation of most magnetic Ap stars, relative to their non-magnetic
brethren, is understood as a result of magnetic braking by the field
threading the stars, through either magnetic coupling to a stellar wind or
``accretion braking'' (Mestel 1975).  That not all A stars show
magnetism is taken to be the result of the different initial conditions
under which the stars formed.  Probably the most pressing question
concerning this theory is whether the primeval field can survive through
the convective Hayashi phase of such stars' pre-main-sequence evolution.
The Hayashi convection may expell the magnetic field from the outer layers
of such a star, perhaps concentrating it in the initially radiative core
(which forms rapidly during the star's descent of the Hayashi track).
Alternatively, a sufficiently strong, concentrated field may be able to
resist expulsion by the convection, and later yield the observed
global-scale fields (e.g., Moss 2001).  A variant of the fossil theory
suggests that the fields were generated by dynamo action driven by the
Hayashi convection, but are not presently being actively maintained against
ohmic decay.

The second approach suggests that the surface magnetic fields may result
from contemporary dynamo activity (e.g., Krause \& Oetken 1976).  A-type
stars possess convective cores, surrounded by extensive envelopes that are
radiative but for very thin shells of convection near the surface.
Convection within the highly conductive plasma of the core, coupled with
rotation, may serve to build strong magnetic fields.  Yet those fields may
well be forever buried from view: diffusion of the fields through the
radiative envelope is thought simply to take too long.  If the
dynamo-generated fields are sufficiently strong, however, they may become
subject to magnetic buoyancy instabilities that could allow them to rise to
the surface where they could be observed (e.g., Moss 1989).  Recent
modeling (MacGregor \& Cassinelli 2003) has provided tantalizing
indications that this process might indeed be able to bring very strong
fibril fields to the surface in a fraction of an A-star's main-sequence lifetime.
However, MacDonald \& Mullan (2004) point out that realistic compositional
gradients slow the rise of such buoyant flux tubes considerably.
Whether the fields built by possible dynamo action within the core are
actually strong enough for such buoyancy instabilities to play a role, or
are instead likely to remain hidden, is thus one of the most pertinent
questions regarding the dynamo approach to explain the surface fields.

Recently an alternative third explanation has emerged, which relies
on the possibility that a radiative envelope could generate mean magnetic field
via dynamo action involving shear layers and the
instability of a large-scale mean toroidal field (Spruit 2002, MacDonald
\& Mullan 2004).

\subsection{Aspects of Core Convection}

Within the cores of A-type stars, the steep temperature gradient that
arises from fusion via the CNO cycle drives vigorous convection.  We have
already examined that convection through extensive hydrodynamic
three-dimensional nonlinear simulations (Browning, Brun \& Toomre 2004,
hereafter BBT; Brun, Browning \& Toomre 2005), in which we solved the
compressible Navier-Stokes equations without magnetism within the anelastic
approximation.  Some of the dynamical properties revealed by such modeling
of rotating convective cores using our anelastic spherical harmonic (ASH)
code are summarized in \S2.4.

In this paper, we turn to magnetohydrodynamic (MHD) simulations of the
dynamo activity that may be occurring within the convective cores of A-type
stars.  Using our prior hydrodynamic simulations as a starting point, we
examine here whether vigorous core convection coupled with rotation can
amplify an initial seed magnetic field and sustain it indefinitely.  Though
we are motivated in part by the remarkable observations of surface
magnetism in Ap stars, the work described here has little to say directly
about such surface fields.  As in the hydrodynamic simulations, we model
only the inner regions of such stars, including the entire convective core
but only a fraction of the overlying radiative zone.  Our principal aim is
simply to explore whether dynamo action occurs at all within such cores
(Browning, Brun \& Toomre 2005), and if so, what are the main properties of
the resulting magnetic fields: their strength, their topology, and their
variability.  Although our work is thus quite preliminary, it should serve
to illuminate some of the complex dynamical processes occurring within Ap
stars.

In \S$2$ we describe our formulation of the problem and briefly summarize
the computational techniques used to address it.  In \S$3$ we summarize the
flows and magnetic fields realized by dynamo action in our simulations, and
consider their evolution with time.  In \S$4$ we examine the mean flows and
transports of angular momentum and heat, and in \S$5$ the many spatial
scales and the spectral distributions of the flows and fields.  In \S$6$ we
consider the evolution of the global-scale axisymmetric poloidal and
toroidal magnetic fields, and in \S$7$ briefly discuss the processes by
which the magnetic fields are generated and sustained. We reflect on the
main findings of this work and their implications in \S$8$.

\section{FORMULATING THE PROBLEM}

\subsection{Convective Core and Radiative Shell}

The simulations considered here are intended to be simplified descriptions
of the inner 30\% by radius of main sequence A-type stars of 2 solar
masses, consisting of the convective core (approximately the inner 15\% of
the star) and a portion of the overlying radiative zone.  Contact is made
with a 1--D stellar model (at an age of 500 Myr) for the initial
conditions, with realistic values for the radiative opacity, density, and
temperature adopted.  We have softened the steep entropy gradient contrast
encountered in going from the convective core to the surrounding radiative
zone, which would otherwise favor the driving of small-scale,
high-frequency internal gravity waves that we cannot resolve with
reasonable computational resources.  This lessened `stiffness' of
the system has some impact on the extent to which convective motions may
overshoot into the radiative region (see BBT).
The inner 2\% of the star is excluded from our computational domain,
for the coordinate systems employed in ASH possess both coordinate
singularities at $r=0$ and decreasing mesh size (and thus quite limited
time steps) with decreasing radius.  Though the exclusion of this central
region might in principle give rise to some spurious physical responses, by
projecting Taylor columns aligned with the rotation axis (e.g., Pedlosky
1987) or by giving rise to boundary layers, we have seen no evidence of
such effects in our simulations.  In trial computations with both smaller
and larger excluded central regions, the developed mean flows were very
similar to those described here.

The main parameters of our simulations are summarized in Table 1.  These
calculations with magnetism were begun by introducing small-amplitude seed
magnetic fields into two statistically mature hydrodynamic simulations from
BBT.  We then followed the evolution of those fields over multiple ohmic
diffusion times.  We have adopted here a magnetic Prandtl number $P_m = 5$,
though $P_m$ in the interiors of real A-type stars is close to unity, which
allows us to achieve higher magnetic Reynolds numbers $R_m$ at moderate
resolution than would be attainable with lower $P_m$.  A detailed
description of the initial conditions and simulation parameters adopted in
our modeling of A star core convection are provided in BBT. Our simulations
are the magnetic analogues of cases {\sl E} and {\sl C4} in that paper,
using these as initial conditions, and denoting the resulting models as
{\sl Em} and {\sl C4m}.  Thus we consider here the central regions of
2-solar mass A-type stars at rotation rates of one and four times the solar
mean angular velocity of {\bf $\Omega_{o}$}$= 2.6\times 10^{-6}$ $s^{-1}$
$= 414$ nHz, corresponding to rotation periods of 28 and 7 days.  Rotation
acts to stabilize these systems against convection (Chandrasekhar 1961), so
our more rapidly rotating case {\sl C4m} was evolved at somewhat lower
values of viscosity and diffusivity than case {\sl Em} rotating at the
solar rate.  Cases {\sl Em} and {\sl C4m} involve different values of the
maximum entropy gradient $d \sb /dr$ in the radiative region, thus sampling
the effects upon penetration as the stiffness of the boundary between that
region and the convective core is varied.

\subsection{Anelastic MHD Equations}
Our ASH code solves the three-dimensional MHD anelastic
equations of motion in a rotating spherical geometry using a pseudospectral
semi-implicit approach (e.g., Clune et al. 1999; Miesch et al. 2000; Brun, Miesch
\& Toomre 2004).  These equations are fully nonlinear in velocity and
magnetic fields and linearized in thermodynamic variables with respect to a
spherically symmetric mean state that is also allowed to evolve.  We take
this spherical mean state to have density
$\bar{\rho}$, pressure $\bar{P}$, temperature $\bar{T}$, specific entropy
$\bar{S}$; perturbations are denoted as $\rho$, $P$,
$T$, and $S$.  The equations being solved are
\begin{eqnarray}
\nab\cdot(\rb {\bf v}) &=& 0, \\
\nab\cdot {\bf B} &=& 0, \\
\rb \left(\frac{\p {\bf v}}{\p t}+({\bf v}\cdot\nab){\bf v}+2{\bf \Omega_o}\times{\bf v}\right) 
 &=& -\nab P + \rho {\bf g} + \frac{1}{4\pi} (\nab\times{\bf B})\times{\bf B} \nonumber \\
&-& \nab\cdot\mbox{\boldmath $\cal D$}-[\nab\bar{P}-\rb{\bf g}], \\
\rb \tb \frac{\p S}{\p t}+\rb \tb{\bf v}\cdot\nab (\sb+S)&=&\nab\cdot[\kappa_r \rb c_p \nab (\tb+T)
+\kappa \rb \tb \nab (\sb+S)]\nonumber \\
&+&\frac{4\pi\eta}{c^2}{\bf j}^2+2\rb\nu\left[e_{ij}e_{ij}-1/3(\nab\cdot{\bf v})^2\right]
 + \rb {\epsilon},\\
\frac{\p {\bf B}}{\p t}&=&\nab\times({\bf v}\times{\bf B})-\nab\times(\eta\nab\times{\bf B}),
\end{eqnarray}
where ${\bf v}=(v_r,v_{\theta},v_{\phi})$ is the velocity in spherical coordinates in 
the frame rotating at constant angular velocity ${\bf \Omega_o}$, ${\bf g}$ is the 
gravitational acceleration, ${\bf B}=(B_r,B_{\theta},B_{\phi})$ is the magnetic field, 
${\bf j}=c/4\pi\, (\nab\times{\bf B})$ is the current density, 
$c_p$ is the specific heat at constant pressure, $\kappa_r$ is the radiative diffusivity, $\eta$ is the 
effective magnetic diffusivity, and ${\bf \cal D}$ is the viscous stress tensor, with components
\begin{eqnarray}
{\cal D}_{ij}=-2\rb\nu[e_{ij}-1/3(\nab\cdot{\bf v})\delta_{ij}],
\end{eqnarray}
where $e_{ij}$ is the strain rate tensor, and $\nu$ and $\kappa$ are
effective eddy diffusivities.  A volume heating term $\rb \epsilon$ is
included in these equations to represent energy generation by nuclear
burning of the CNO cycle within the convective core.  To close the set of
equations, the thermodynamic fluctuations are taken to satisfy the linearized relations
\begin{equation}\label{eos}
\frac{\rho}{\rb}=\frac{P}{\bar{P}}-\frac{T}{\tb}=\frac{P}{\gamma\bar{P}}
-\frac{S}{c_p},
\end{equation}
assuming the ideal gas law
\begin{equation}\label{eqn: gp}
\bar{P}={\cal R} \rb \tb ,
\end{equation}

\noindent where ${\cal R}$ is the gas constant.  The effects of
compressibility on the convection are taken into account by means of the
anelastic approximation, which filters out sound waves that would otherwise
severely limit the time steps allowed by the simulation.  In the MHD
context here, the anelastic approximation filters out fast magneto-acoustic
modes but retains the Alfven and slow magneto-acoustic modes.  In order to ensure
that the mass flux and the magnetic field remain divergence-free to machine
precision throughout the simulation, we use a toroidal--poloidal
decomposition 
\begin{eqnarray}
{\rb\bf v}=\nab\times\nab\times (W {\bf e}_r) +  \nab\times (Z {\bf
  e}_r) , \\ 
{\bf B}=\nab\times\nab\times (C {\bf e}_r) +  \nab\times (A {\bf e}_r) ~~~, 
\end{eqnarray}
with ${\bf e}$ a unit vector, and involving the streamfunctions $W$,
$Z$ and magnetic potentials $C$, $A$,
which are functions of all three spatial coordinates plus time.  

The full set of anelastic MHD equations solved by ASH is described in Brun,
Miesch \& Toomre (2004), though dealing there with solar dynamo processes in a deep
convective shell.  In order to be well-posed, our system of equations for
$W$, $Z$, $C$, and $A$, and for the fluctuating entropy $S$ and pressure $P$, 
requires 12 boundary conditions and suitable initial
conditions. Since we aim to assess the angular momentum redistribution in
our simulations, we have opted for torque-free velocity and magnetic
boundary conditions at the top and bottom of the deep spherical domain.
These are symbolically

 $a$. impenetrable top and bottom surfaces: $v_r=0|_{r=r_{bot},r_{top}}$,

 $b$. stress free top and bottom: $\frac{\p}{\p
 r}\left(\frac{v_{\theta}}{r}\right)=\frac{\p}{\p
 r}\left(\frac{v_{\phi}}{r}\right)=0|_{r=r_{bot},r_{top}}$,

 $c$. constant entropy gradient at top and bottom: $\frac{\p \sb}{\p
r}=$ constant$|_{r=r_{bot},r_{top}}$,

 $d$. purely radial magnetic field at top and bottom (match to a highly permeable 
external media, Jackson 1999) : $B_{\theta}=B_{\phi}=0|_{r=r_{bot},r_{top}}$.  

\noindent Requiring the magnetic field to be purely radial at the boundaries means
that the Poynting flux vanishes there, with no magnetic energy leaking
out of the domain.

\subsection{Numerical Approach}

Convection in stars occurs on many spatial scales.  No numerical
simulations can presently consider all these scales simultaneously.  We
choose to resolve the largest scales of the nonlinear flows and magnetic
fields, which we think are likely to be the dominant players in
establishing differential rotation and other mean properties of the core
convection.  Our large-eddy simulations (LES) thus explicitly follow the
larger scales, while employing sub-grid-scale (SGS) descriptions of the
effects of unresolved motions.  Those unresolved motions are manifested
simply as enhancements to the kinematic viscosity and thermal and magnetic
diffusivities ($\nu$, $\kappa$, and $\eta$ respectively), which are thus
effective eddy viscosities and diffusivities.  For simplicity, we have
taken these to be functions of radius alone, and chosen to scale them as the inverse of
the square root of the mean density.  We are encouraged by the relative
successes that similar simulations (e.g., Miesch et al. 2000; Elliott et
al. 2000; Brun \& Toomre 2002) have achieved in matching the detailed
observational constraints provided by helioseismology on differential
rotation achieved by solar convection.  However, we recognize that
considerable refinements for SGS treatments are generally needed, and such
work is under way.  

Within ASH, the dynamic variables are expanded in spherical harmonics
$Y^m_{\ell}(\theta,\phi)$ in the horizontal directions and in Chebyshev
polynomials $T_n (r)$ in the radial. Thus spatial resolution is uniform
everywhere on a sphere when a complete set of spherical harmonics of degree
$\ell$ is used, retaining all azimuthal orders $m$ in what is known as a
triangular truncation.  We here limit our expansion to degree
$\ell=\ell_{max}$, which is related to the number of latitudinal mesh
points $N_{\theta}$ (here $\ell_{max}=(2N_{\theta}-1)/3$), take $N_{\phi}=2
N_{\theta}$ latitudinal mesh points, and utilize $N_r$ collocation points
for the projection onto the Chebyshev polynomials.  We employ a stacked
Chebyshev representation, wherein the computational domain is split into
two regions and separate Chebyshev expansions performed for each.  We thus
attain higher resolution at the interface between these two regions, here
set as the approximate boundary between the convective and radiative zones,
in order to capture better the penetrative convection occurring there.
We have taken $N_r=49+33=82$ and $\ell_{max}=170$ in the simulations
considered here.  The time evolution of the linear terms is determined
using an implicit, second-order Crank-Nicholson scheme, whereas an explicit
second-order Adams-Bashforth scheme is employed for the advective, Lorentz,
and Coriolis terms.  The ASH code has been optimized to run efficiently on
massively parallel supercomputers such as the IBM SP-4 and the Compaq TCS-1
using the message passing interface (MPI), and has demonstrated good
scalability on such machines up to about 1000 processors.  More details on
the numerical implementation of ASH are provided in Clune et al. (1999) and
in Brun et al. (2004).

The intricate and sustained time dependence typical of core convection
requires extended simulation runs to assess the dynamical equilibration of
such systems, spanning over 7000 days of physical time (or about 300
rotation periods) in one of our cases.  The analysis of such dynamics
requires forming various spatial and temporal averages of the evolving
solutions.  We will use the symbol $\hat{a}$ to indicate temporal and
longitudinal averaging of say the variable $a$, and the symbol $<a>$ in
denoting longitudinal averaging alone to obtain the axisymmetric component
of the variable.  The latter allows us to separate the fluctuating (denoted
by the prime as $a'$) from the axisymmetric (mean) parts of the variable.
This is convenient, for instance, in defining fluctuating and mean velocity
components (relative to the rotating frame).  The symbol $\tilde{a}$
designates the rms average of the variable, carried out over a spherical
surface for many realizations in time.  Likewise, the combined symbols
$\tilde{a}'$ represent similar rms averaging of the variable from which the
axisymmetric portion has been subtracted.

\subsection{Progenitor Convection with Differential Rotation}

\begin{figure}[hpt]
\center
\epsscale{1.0}
\plotone{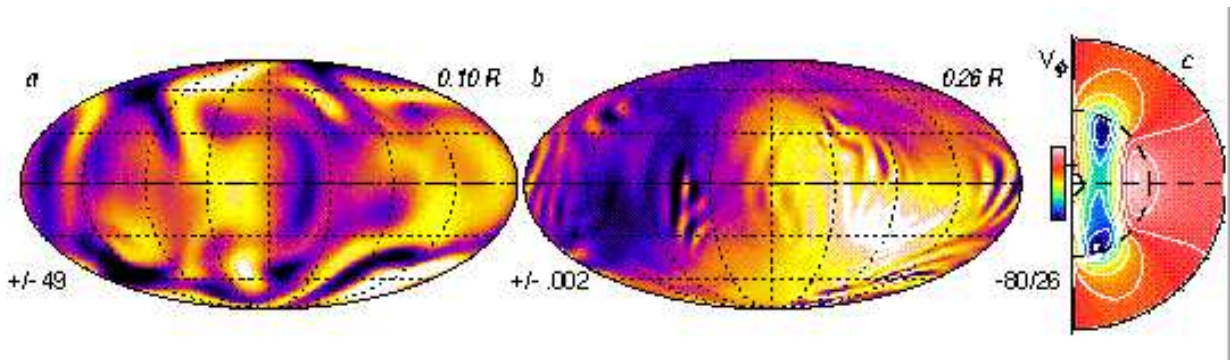}
\caption{\label{hydro} Flow properties of core convection in the progenitor hydrodynamic
simulation case {\sl E} from BBT.  $(a)$ Radial velocity
$v_r$ at one instant in mid-core (at $r=0.10 R$), shown in a global view as
a Mollweide projection.  Broad regions of upflow are in light tones, and
downflows dark, as indicated by color bar with ranges in m s$^{-1}$.  $(b)$
Companion view of $v_r$ within surrounding radiative envelope (at $r=0.26
R$), showing signature of the relatively weak internal gravity waves
excited by the plumes of penetration.  $(c)$ Resulting differential
rotation established in the computational domain displayed in cross-section
of radius and latitude.  Shown as a contour plot is the time and longitudinally averaged
zonal velocity $\vph$, which possesses a central column of
particularly slow rotation (retrograde relative to the frame).  The equator
is denoted by the dashed line, the rotation axis is vertical, and the outer
extent of the prolate core is indicated by the dotted curve.}
\end{figure}

The simulations described here take as their starting points an evolved
instant in the hydrodynamic simulations of core convection described in
BBT. We illustrate in Figure \ref{hydro} some of the striking dynamical
properties revealed by one of those progenitor simulations.  Figure
\ref{hydro}$a$ shows a global mapping at one instant of the radial velocity
deep within the core in case {\sl E}, which is rotating at the solar rate.
In this Mollweide projection, meridian lines are seen as curved arcs, and
lines of constant latitude are indeed parallel.  Convection within the core
involves broad sweeping flows that span multiple scale heights, with little
apparent asymmetry between upflows (light features) and downflows (dark
tones).  The convective flows in such global domains can readily plunge
through the center, thus coupling widely separated sites.  The flows are
highly time dependent, with complex and intermittent features emerging as
the simulations evolve.  Such vigorous convection is able to penetrate into
the overlying radiative zone, with that extent varying with latitude.  The
upward-directed penetrating plumes serve to excite gravity waves in the
stable envelope, seen in Figure \ref{hydro}$b$ as localized ripples on many
scales.

The coupling of convection with rotation in these spherical geometries
yields a prominent differential rotation exhibited in Figure
\ref{hydro}$c$.  The mean zonal flows shown there (relative to the rotating
frame) are characterized by a central cylindrical column of slow rotation.
Within the bulk of the convection zone, this differential rotation is
driven primarily by the Reynolds stresses associated with the convection,
helped by meridional circulation and opposed by viscous stresses.  
Near the interface between the core and the radiative envelope, baroclinicity 
also plays an important role.

The penetrative convection yields a nearly adiabatically stratified core
region that is prolate in shape and aligned with the rotation axis (shown
by the dashed curve in Figure \ref{hydro}$c$).  This is surrounded by a
further region of overshooting in which the convective plumes can mix the
chemical composition but do not appreciably modify the stable
(subadiabatic) stratification.  The outward extent of this zone is roughly
spherical.  Our progenitor simulations in BBT have thus revealed that core
convection establishes angular velocity profiles with a distinctive central
column of slowness, a prolate shape to the well-mixed core, and a broad
spectrum of gravity waves in the radiative envelope.

\section{DYNAMO ACTION REALIZED IN CORE}

We have found that vigorous core convection coupled with rotation clearly
admits magnetic dynamo action.  The initial seed magnetic fields introduced
into our two progenitor hydrodynamic simulations are amplified greatly by
the convective and zonal flows, ultimately yielding magnetic fields that
possess energy densities comparable to that in the convection itself.  Here
we begin by assessing the growth of the magnetic fields and their saturation, the
morphology of the magnetism and the resulting modified convection, and the
intricate time dependence of the sustained fields and flows.

\subsection{Growth and Saturation of Magnetic Fields}

The temporal evolution of the magnetic energy (ME) and kinetic energy (KE)
densities (volume-integrated and relative to the rotating frame) in case
{\sl C4m} is displayed in Figure \ref{timetrace}$a$.  The magnetic field
undergoes an initial phase of exponential growth from its very weak seed
field that lasts about 1000 days.  In case {\sl Em} (not shown), the
initial phase of growth lasts about 1700 days. The seed dipole fields are
in both simulations amplified by more than eight orders of magnitude. The
different growth rates for the magnetic field realized in the two
simulations result from the differing Reynolds numbers and magnetic
diffusivities adopted. Both have magnetic Reynolds numbers (see Table 1)
well in excess of the threshold values that earlier studies of convection
in spherical shells (e.g., Gilman 1983; Brun et al. 2004) have found
necessary for dynamo action, typically $R_m \approx 300$.  This exponential
growth is followed by a nonlinear saturation phase, during which the
Lorentz forces acting upon the flows yield statistical equilibria in which
induction is balanced in the large by ohmic dissipation.  The magnetic
field attains different saturation amplitudes in the two simulations.  In
case {\sl C4m}, the energy density in the magnetic field (ME) is typically
about $88$\% of the kinetic energy density (KE), whereas in case {\sl Em}
it is about 28\% but fluctuates considerably in several phases of
behavior. Such field amplitudes are sustained for longer than the magnetic
diffusion time across the computational domain -- here, $\tau \sim
L^2/(\pi^2 \eta) \sim 3900$ days (Moffatt 1978) -- implying that the
magnetic field is being actively maintained against ohmic decay.  Thus
sustained dynamo action has probably been realized.  The different set of
parameters used in the two simulations appear to account for the different
saturation field strengths that are realized.

The strong magnetic fields established by dynamo action within the core are
expected to interact with the convective and zonal flows.  A clear
indication of such feedback is provided by the reduction of KE visible in
Figure \ref{timetrace}$a$ after about $1000$ days.  This first becomes
apparent once ME grows to about 1\% of KE, as the magnetic fields begin to
significantly modify the flows through the Lorentz (${\bf j} \cross {\bf B}$) term in equation
(3), much as in simulations of strong dynamo activity in the solar
convection zone (Brun, Miesch \& Toomre 2004).  The reduction in KE here is due
primarily to a significant decline in the energy contained in the
differential rotation (DRKE).  In simulation {\sl C4m}, DRKE decreases to
only 3\% of its value in the hydrodynamic progenitor simulation {\sl C4}
(see also Table 3). In case {\sl Em}, the decline is also appreciable, with
DRKE dropping to 19\% of its value in the hydrodynamic simulation. We
consider issues of the resulting differential rotation and its linkage to
temporal variations of ME in \S4.1.

Like the convective and zonal flows that build and sustain them, the
magnetic fields in these simulations are highly variable in time.  This
variability is apparent in Figure \ref{timetrace}$b$, which
shows the evolution of various energy densities over an interval subsequent
to the initial exponential growth of the magnetic field.  Shown are the
energies in the convection (CKE) together with KE, DRKE, and ME. Though no
continuous growth or decline of the energy densities is evident, they show
considerable variations for this case {\sl C4m}.  During this interval, KE
fluctuates by about a factor of three, with most of this variation
reflecting that of CKE. The modulations in CKE have a temporal spacing of
about 130-140 days, or roughly 20 rotation periods. Here ME likewise varies, as the convective and
zonal flows serve to modify the magnetic fields through the production term
in the induction equation (5). Indeed, during some intervals in the
evolution of case {\sl C4m} shown in Figure \ref{timetrace}$b$, ME actually
exceeds KE.  It is interesting that the field strengths achieved in case
{\sl C4m} thus roughly represent equipartition between the flows relative
to the rotating frame and the magnetism.  Such values of ME represent
typical rms field strengths in the core of about 67 kG, as compared to rms
flow velocities that are about 30 m s$^{-1}$ (Table 2).

\subsection{Morphology of Flows and Magnetism}

Within the core, broad convective flows sweep through the spherical domain,
with large-scale regions of upflow and downflow serving to couple widely
separated regions.  The global connectivity permitted in these full
spheres, together with the fairly small density contrasts present, results
in motions that can span large fractions of a hemisphere and extend
radially through much of the convection zone.

Such global-scale convective flows are apparent in Figure \ref{hvr}$a$,
which shows a volume rendering of a snapshot of the radial velocity $v_r$
near the outer boundary of the convective core in case {\sl C4m}.  The
region of vigorous convection is slightly prolate in shape, much as in the
progenitor, extending farther in radius near the poles than near the
equator.  No obvious asymmetry between regions of upflow and downflow is
visible.  This stands in sharp contrast to the results of solar convection
simulations that exhibit broad upflows together with narrow and fast
downflows.

The magnetic fields sustained within the convection zone are characterized
by smaller scale features than are present in the convective flows.  The
intricate nature of the field is most apparent in Figure \ref{hvr}$b$,
which shows the radial component of magnetic field $B_r$.  Here the field appears as a
tangled collection of positive and negative polarity on many different
scales.  The finer structure present in the magnetic fields than in the
convective flows comes about partly because we have taken the magnetic
diffusivity to be smaller than the viscous diffusivity (with $P_m = 5$).

The longitudinal fields $B_{\phi}$ shown in Figure \ref{hvr}$c$ likewise
possess small-scale structure, but also exhibit organized bands of
magnetism that wrap around much of the core.  These broad ribbons of
toroidal field may arise due to stretching by gradients of angular velocity
near the interface between the core and the radiative envelope.  Such
stretching and amplification of toroidal field by differential rotation,
described in mean-field theories as the $\omega$-effect, mirrors what is
thought to occur in the tachocline of rotational shear at the base of the
solar convection zone.  In the sun, magnetic fields are thought to be
pumped downward from the envelope convection zone into the radiative interior, with
the tachocline at the interface producing strong toroidal fields that eventually rise by
magnetic buoyancy through the convection zone (e.g., Charbonneau \&
MacGregor 1997).  Here we may be seeing the reverse analog of such a process in
stars with convective interiors surrounded by radiative envelopes.

The intricate networks of magnetic fields and convective flows are also
revealed in Figure \ref{vrbrbp_Em} (for case {\sl Em}) and 
in Figure \ref{vrbrbp_C4m} (for case {\sl C4m}) by global mappings of the radial velocity ($v_r$)
together with the radial ($B_r$) and azimuthal ($B_{\phi}$) magnetic fields
at two depths.  Deep within the convective core (at $r=0.10 R$, \emph{right}), 
the finely threaded magnetic field coexists with the relatively
broad patchwork of convective flows.  Features in $v_r$ and $B_r$ possess evident
links, with the convective downflow lanes containing strong radial magnetic
field of both polarities.  In contrast, upflows contain few strong magnetic
structures.  The azimuthal field $B_{\phi}$ within the core appears to be
quite patchy, with little correlation to the radial velocity field $v_r$.
Finer structure is present in all the fields displayed for case {\sl C4m}
(Fig. \ref{vrbrbp_C4m}), relative to case {\sl Em}, owing mostly to the
slightly smaller viscosities and resistivities adopted for {\sl C4m}.

Near the boundary of the convective core and the radiative envelope
($r=0.16 R$, \emph{left}), $v_r$ and $B_r$ both possess considerably
smaller amplitudes than in the deep interior, with $v_r$ in case {\sl Em}
(Fig. \ref{vrbrbp_Em}) lessened by a factor of 240 and $B_r$ by a factor of
100.  This suggests that the spherical surface shown cuts through a region
where only weak overshooting of the convection survives.  In case {\sl
C4m} (Fig. \ref{vrbrbp_C4m}$a$,$b$), the amplitudes of $v_r$ and $B_r$ are
reduced by smaller factors (of 64 and 18 respectively) in going from
$r=0.10R$ to $r=0.16R$, most likely because one is sampling here the
penetrative convection more directly, probably because of the weaker stable
stratification in this case.  In contrast, $B_{\phi}$ near the core
boundary in both cases is only slightly diminished from its interior
strength, and may reflect the continuing production of toroidal field there
by rotational shear.  The overall magnetic fields at the core boundary are
of larger physical scale than the fields deeper down, and $B_{\phi}$
(Figs. \ref{vrbrbp_Em}$c$, \ref{vrbrbp_C4m}$c$) shows the same broad and
wavy, ribbon-like features evident in the volume renderings of Figure
\ref{hvr}$c$. In case {\sl C4m} (Fig. \ref{vrbrbp_C4m}$c$), the magnetic
field at this radius ($r=0.16 R$) is much stronger at high latitudes than
at the equator, reflecting the prolate shape of the strongly magnetized
core of convection.  The spherical surface viewed here lies inside this
prolate region near the poles, but outside it at the equator.  Thus the
stronger influence of rotation in this case {\sl C4m} has yielded greater
departures from a spherical shape for the core with penetration than is
realized in case {\sl Em}, helped also by the reduced stiffness of the
radiative envelope in case {\sl C4m}.

Our global mappings (at $r=0.16 R$) also reveal that the pummelling of the base of the
radiative envelope by the upward-directed convective plumes serves to
excite a broad range of internal gravity waves.  These waves are visible at
the low latitudes in Figures \ref{vrbrbp_Em}$a$, \ref{vrbrbp_C4m}$a$ as
low-amplitude ripples of small physical scales.  Similar gravity waves were
seen in the progenitor non-magnetic simulations in BBT.

\subsection{Time-Dependence of Sustained Flows and Fields}

The convective flows and the magnetic fields that they generate in our two
cases evolve in a complicated fashion.  Throughout the convective core, we
have observed the birth of magnetic structures, their advection and
shearing by the flows, and their mergers with other features or cleaving
into separate structures.  Some flows and magnetic structures persist for
many days, while others rapidly fade away.  A brief sampling of such
behavior in case {\sl Em} is provided in Figure \ref{timevol_Em} showing a
succession of spherical views of both $v_r$ and $B_r$ in mid-core ($r=0.10
R$) at four closely spaced snapshots (each 6 days apart). Several features
amidst the magnetism, labeled $A$, $B$, $C$ in Figure \ref{timevol_Em},
propagate in a slightly retrograde fashion (to the left) over the interval
sampled.  Features $A$ and $B$ remain confined to low latitudes, with
feature $B$ varying considerably in strength and size as the simulation
evolves.  In Figure \ref{timevol_Em}$e$, this structure is visible as a
weak patch of negative $B_r$ at a latitude of about 20\degr; later it has
become a much broader feature of greater amplitude
(Fig. \ref{timevol_Em}$h$), seen as a dark patch at a latitude of about
15\degr.  Feature $C$, which at first appears as a narrow structure of
negative polarity spanning latitudes from the equator to about 45\degr,
then propagates toward higher latitudes, and is sheared and weakened.  The
convective flows exhibit similar changes, with a coherent downflow lane
(Fig. \ref{timevol_Em}$a$) spanning both hemispheres gradually breaking up
into multiple structures.

Such rich time dependence is assessed over longer temporal intervals in
case {\sl Em} by turning to time-longitude mappings in Figures
\ref{timelongEm_eq} and \ref{timelongEm_60}. These show the variation with
time of $v_r$ and $B_r$ sampled (at $r=0.10 R$) for all longitudes either
at the equator (Fig. \ref{timelongEm_eq}) or at 60$\degr$\ latitude
(Fig. \ref{timelongEm_60}).  Coherent downflow lanes are visible in these
time-longitude mappings of $v_r$ as dark bands tilted to the right,
indicating prograde propagation (relative to the frame), or to the left or
retrograde, which often persist for multiple rotation periods.  Similar
evolution is observed in the companion mappings of $B_r$, with most major
structures evident in both the flows and the magnetism. Much as in Figure
\ref{vrbrbp_Em}, the persistent convective downflows contain magnetic
fields of mixed polarities, whereas the upflows are largely devoid of
strong magnetic structures.  The propagation of these large-scale
structures tends to be prograde at the equator (Fig. \ref{timelongEm_eq})
and strongly retrograde at high latitudes (Fig \ref{timelongEm_60}).  There
is also substantial evolution of the flows on short time scales, with some
striking features of the convection rapidly emerging and then fading.
Similarly, the magnetic field exhibits both rapid evolution of some 
structures, and others that survive for extended periods of time.  Identification of
persistent features amidst the magnetism is occasionally made more
difficult by the finely threaded field topology. However, structures
evident in $B_r$ generally appear to be advected and to propagate in
roughly the same fashion as features in $v_r$, with both tending to wax and
wane as the simulations evolve.

\section{MEAN FLOWS AND TRANSPORT}

In the deep spherical domains studied here, the Coriolis forces associated
with rotation can have major impacts on the structure of the convective
flows and thus on the manner in which they redistribute angular momentum.
When that influence is strong, as when the convective overturning time is
at least as long as the rotation period (with the convective Rossby number
$R_{c}$ of order unity or smaller), a strong differential rotation may be
achieved and maintained.  This was realized in all the cases studied in
BBT.  The dynamo action and consequent intense magnetic fields realized in
our current simulations must feed back strongly on the convection through
the Lorentz forces, probably reducing the differential rotation that can be
maintained.  Intuitively, one expects that the presence of magnetic fields
will tend to diminish the differential rotation, with the field lines that
thread the core acting like rubber bands to couple disparate regions and
enforce more uniform rotation.  Such an analogy is too simple given the
tangled and intermittent nature of the magnetic fields in our simulations,
yet the expectation that the presence of magnetism leads to reduced
differential rotation turns out to be largely correct.  We now consider the
mean zonal flows of differential rotation that are realized in our cases
{\sl Em} and {\sl C4m}, their variations in time, and the manner in which
they are sustained.

\subsection{Nature of Accompanying Differential Rotation}

The differential rotation profiles achieved in our two cases {\sl Em} and
{\sl C4m} are shown in Figure \ref{diffrotn_maghydro}.  These are displayed
first as contour plots with radius and latitude of the longitudinal (or
zonal) velocity $\vph$, with the hat denoting an average in time and longitude.  Shown
also are plots of the radial variation of the associated angular velocity
$\hat{\Omega}$ along three latitudinal cuts, contrasting the behavior in
our magnetic simulations with that of their progenitors.  The latter
emphasize that in case {\sl Em} the angular velocity contrasts have been
lessened almost twofold from the hydrodynamic progenitor.  In case {\sl
C4m} that contrast has been nearly eliminated.  The contours of
$\vph$ emphasize that central columns of slow rotation are
realized in both cases, as in their progenitors, but with reduced zonal
flow amplitudes (see Fig. \ref{hydro}$c$).  Both $\vph$ profiles exhibit
some asymmetry between the northern and southern hemisphere, with such
behavior more pronounced for case {\sl Em}
(Fig. \ref{diffrotn_maghydro}$a$).  Noteworthy for case {\sl C4m} is that
the column of slowness in $\vph$ extends well into the radiative envelope,
owing in part to the stronger meridional circulations exterior to the core
achieved with the faster frame rotation.  The longitudinal velocity in {\sl
C4m} appears to be nearly constant on cylinders aligned with the rotation
axis, somewhat akin to Taylor-Proudman columns achieved when rotational
constraints are strong.

With the temporal changes seen in our convective flows and magnetism
(Fig. \ref{timevol_Em}) come also substantial variations in the
differential rotation that they establish.  This is most pronounced in case
{\sl Em}.  Figure \ref{timetrace_Em} shows a detailed view of temporal
fluctuations in volume-averaged energy densities of the differential
rotation (DRKE), convection (CKE), and magnetism (ME).  These reveal that
during extended intervals DRKE exceeds ME, but with moderate oscillations;
such an interval was sampled in producing Figure \ref{diffrotn_maghydro}.
The fairly regular accompanying oscillations in CKE (and thus also KE) have
periods of about 150-200 days, as contrasted to the rotation period of 28
days.  There are also remarkable brief intervals during which DRKE plummets
by nearly an order of magnitude, with two such shown in Figure
\ref{timetrace_Em} at about 4200 days and 7000 days in the simulation.  The
onset of those grand minima in DRKE coincide with times when ME has climbed
to values greater than about 40\% of KE.  This suggests that strengthening
magnetic fields can lead to abrupt collapses in the differential rotation
established by the convection, followed by a recovery.  This arises partly
from the strong feedback of the Lorentz forces on the convection and on the
differential rotation, both of which serve to build the fields through
induction.  With the consequent diminished flows, the field production is
lessened, and so the magnetic fields weaken. Once below a given threshold
(here ME less than 40\% of KE), the convection regains its strength,
leading to stronger Reynolds stresses (see \S 4.2) that reestablish the
differential rotation, with magnetic induction once again invigorated.
Thus the cycle lasting about 2000 days begins anew.  Such intricate
behavior seen in case {\sl Em} is not realized in case {\sl C4m} where ME
and KE are always comparable though moderately variable (see
Fig. \ref{timetrace}$b$).  Since ME in this case is far stronger, a cyclic
behavior in which the Lorentz forces oscillate between being strong or weak
is not realized. The complex changes in the differential rotation achieved
in case {\sl Em} are shown in Figure \ref{3omegas} that samples three short
temporal averages of $\vph$ and $\hat{\Omega}$.  These examine intervals
prior, during, and after the second pronounced minimum of DRKE
(Fig. \ref{timetrace_Em}).  During that minimum the angular velocity
contrast within the core (Fig. \ref{3omegas}$b$) is modest, and the
retrograde column of slowness in $\vph$ is barely there.  The samples
before (Fig. \ref{3omegas}$a$) and after (Fig. \ref{3omegas}$c$) show zonal
flows and angular velocity contrasts much as in Figures
\ref{diffrotn_maghydro}$a$,$b$, possessing central regions of slow
rotation.  In contrast, no comparable large variation in zonal flows are
realized in case {\sl C4m}, where angular velocity contrasts are modest at
all times, much as in the long time average shown in Figure
\ref{diffrotn_maghydro}$d$,$e$.

\subsection{Redistributing the Angular Momentum}

The complex MHD systems studied here exhibit a rich variety of
responses, with intricate time dependence seen in both the flows and magnetic
fields.  How the zonal flows seen as differential rotation arise and are
sustained, how they interact with the magnetism, and how they vary in time
are thus all sensitive matters.  This behavior cannot now be predicted from
first principles, but the present simulations offer a unique opportunity to
determine the roles played by different agents in transporting angular
momentum and giving rise to the differential rotation.  Since our case {\sl
Em} exhibits strong angular velocity contrasts, albeit variable, we will
here examine how these are established.

Our simulations were conducted with stress-free and purely radial magnetic field  
boundary conditions, so no net external torque is applied to the
computational domain.  Thus total angular momentum within the simulations
is conserved.  We can assess the transport of angular momentum within these
systems in the manner of Brun, Miesch \& Toomre (2004) (see also Elliott,
Miesch, \& Toomre 2000).  We consider the $\phi$-component of the momentum
equation expressed in conservative form and averaged in time and longitude:

\begin{equation}
\frac{1}{r^2} \frac{\p(r^2 {\cal F}_r)}{\p r}+\frac{1}{r \sin\theta}
\frac{\p(\sin \theta {\cal F}_{\theta})}{\p
\theta}=0,
\end{equation}
involving the mean radial angular momentum flux
\begin{equation}
{\cal F}_r=\rb r\sin\theta[-\nu r\frac{\p}{\p
r}\left(\frac{\hat{v}_{\phi}}{r}\right)+\widehat{v_{r}^{'}
v_{\phi}^{'}}+\hat{v}_r(\hat{v}_{\phi}+\Omega r\sin\theta)-\frac{1}{4\pi\rb}\widehat{B_{r}^{'}
B_{\phi}^{'}}-\frac{1}{4\pi\rb}\hat{B}_r\hat{B}_{\phi}] \end{equation}
and the mean latitudinal angular momentum flux
\begin{equation}
{\cal F}_{\theta}=\rb r\sin\theta[-\nu
\frac{\sin\theta}{r}\frac{\p}{\p
\theta}\left(\frac{\hat{v}_{\phi}}{\sin\theta}\right)+\widehat
{v_{\theta}^{'} v_{\phi}^{'}}+\hat{v}_{\theta}(\hat{v}_{\phi}+\Omega
r\sin\theta)-\frac{1}{4\pi\rb}\widehat{B_{\theta}^{'}
B_{\phi}^{'}}-\frac{1}{4\pi\rb}\hat{B}_{\theta}\hat{B}_{\phi}].
\end{equation}

In the above expressions, the terms on both
right-hand-sides denote contributions respectively from viscous
diffusion (which we denote as ${\cal F}_r^{VD}$ and ${\cal
F}_\theta^{VD}$), Reynolds stresses (${\cal F}_r^{RS}$ and ${\cal
F}_\theta^{RS}$), meridional circulation (${\cal F}_r^{MC}$ and ${\cal
F}_\theta^{MC}$), Maxwell stresses (${\cal F}_r^{MS}$ and ${\cal
F}_\theta^{MS}$) and large-scale magnetic torques (${\cal F}_r^{MT}$
and ${\cal F}_\theta^{MT}$). The Reynolds stresses are associated with
correlations of the fluctuating velocity components (shown primed) that arise from
organized tilts within the convective structures. Similarly, 
the Maxwell stresses are associated with correlations
of the fluctuating magnetic field components that arise from tilt
and twist within the magnetic structures.

Analyzing the components of  ${\cal F}_r$ and
${\cal F}_{\theta}$ is aided by integrating over co-latitude and radius to deduce the net fluxes through
shells at various radii and through cones at various latitudes, such that
\begin{equation}
I_r(r)=\int_0^{\pi} {\cal F}_r(r,\theta) \, r^2 \sin\theta
\, d\theta \; \mbox{ , } \; I_{\theta}(\theta)=\int_{r_{bot}}^{r_{top}} {\cal
F}_{\theta}(r,\theta) \, r \sin\theta \, dr \, .
\end{equation}
We then identify in turn the contributions from viscous diffusion (VD),
Reynolds stresses (RS), meridional circulation (MC), Maxwell stresses (MS)
and large-scale magnetic torques (MT). This helps to 
assess the sense and amplitude of angular momentum transport within the
convective core and the radiative exterior by each component of ${\cal F}_r$ and ${\cal
F}_{\theta}$.  We now examine the transports achieved within case {\sl Em}
by temporally averaging the fluxes over the interval spanning from 6700 to
7000 days, during which the system was undergoing changes (see
Fig. \ref{timetrace_Em}).  

Turning first to the integrated radial fluxes of angular momentum in Figure
\ref{amom_balance}$a$, we see that the Maxwell stresses ($I_r^{MS}$) are
playing a major role in the radial transport, acting in the outer portions
of the core to transport angular momentum radially outwards, and deeper
down to transport it inwards. In this they are opposed principally by
meridional circulations ($I_r^{MC}$), and aided by the Reynolds stresses
($I_r^{RS}$) associated with the convective flows.  The strong Maxwell
stresses realized in our simulations are noteworthy, for they lead here to
major departures from the angular momentum balance that was achieved in the
progenitor hydrodynamic models.  Over the evolution interval for case {\sl
Em} sampled in Figure \ref{amom_balance}$a$, the Maxwell stresses act in
concert with the Reynolds stresses throughout much of the core, even though
the corresponding terms in equation (12) carry opposite signs.  This
indicates that correlations between the radial and longitudinal components
of the fluctuating magnetic field are reversed with respect to those of the
fluctuating velocity field.  Such behavior was not realized in the solar
convection simulations of Brun et al. (2004), and is less pronounced in our
companion case {\sl C4m}.  We also see that the torques provided by the
axisymmetric magnetic fields ($I_r^{MT}$) are small throughout most of the
convective core, in keeping with the finding in \S6 that the mean
axisymmetric fields are dwarfed in strength by the fluctuating ones.
However, near the outer boundary of the core these mean magnetic torques
grow more significant, in keeping with the mean fields there becoming a
significant contributor to the magnetic energy.  There they act together
with the Maxwell and Reynolds stresses to transport angular momentum
outwards.  The viscous flux is everywhere negative and fairly small
relative to the other components. All of the component fluxes decrease
rapidly outside of the convective core, as both convective motions and
magnetic fields vanish.

The net radial flux $I_r$ (Fig. \ref{amom_balance}$a$) would be zero in a
steady state but here is markedly negative.  However, since in case {\sl
Em} the differential rotation shows prominent changes with time, there must
be non-zero net fluxes of angular momentum to accomplish such changes.
Over the interval sampled by Figure \ref{amom_balance}, the system is
transitioning from a state of high DRKE -- characterized by a strongly
retrograde core -- to one of low DRKE with only small angular velocity
contrasts (see Fig. \ref{timetrace_Em}).  Thus the central regions of the
convective core are being spun up, and so there must be a net angular
momentum flux inward.  Figure \ref{amom_balance}$a$ confirms that this is
indeed occurring during this interval.  

The integrated latitudinal angular momentum fluxes in Figure
\ref{amom_balance}$b$ also reveal a complex interplay among the different
transport mechanisms.  Here the Maxwell stresses ($I_{\theta}^{MS}$) act
largely to slow down the equator (by transporting angular momentum toward
the poles), opposing the strong Reynolds stresses ($I_{\theta}^{RS}$) that
seek to accelerate it.  Thus, in contrast to the radial integrated fluxes,
the Reynolds and Maxwell stresses transport angular momentum in
opposite directions. Similar results for the respective role of the
Reynolds and Maxwell stresses in tranporting angular momentum latitudinally
are found in the solar magnetic cases computed by Brun et al. (2004).
Meridional circulations ($I_{\theta}^{MC}$) also generally act to
accelerate the equator, though the complicated multi-celled nature of those
circulations makes the angular momentum flux they provide decidedly
nonuniform.  The weak axisymmetric magnetic torques ($I_{\theta}^{MT}$),
like their strong fluctuating counterparts the Maxwell stresses, act to
oppose the equatorial acceleration afforded by the Reynolds stresses.
Viscous diffusion plays only a small role, but also tends to transport
angular momentum away from the equator.

Although Figure \ref{amom_balance} assesses the angular momentum transports
during an interesting interval marked by changes in the differential
rotation, the character of the various contributing fluxes is much the same
during other intervals.  Examining these fluxes provides clues as to why
the magnetic simulations exhibit much weaker differential rotation (or
DRKE) than their progenitors.  Whereas in the progenitor the Reynolds
stresses $I_{\theta}^{RS}$ that sought to accelerate the equator competed
only against meridional circulations and viscous diffusion, here they must
also counteract the poleward transport of angular momentum provided by the
Maxwell stresses and large-scale magnetic torques
(Fig. \ref{amom_balance}$b$). Though in principle the fluxes due to the
Reynolds stresses and meridional circulations could adjust to compensate
for such poleward transport, this was not realized in case {\sl Em}.  Thus
the speeding up of the equatorial regions of the outer core was lessened,
and so too the slowing down of the central column, with an overall decrease
in the angular velocity contrast.

\subsection{Radial Transport of Energy}

Since convection in the core arises because of the need to move energy
radially outward, we now assess the role of different agents in transporting the energy within
our simulations. Figure \ref{fluxbalance} presents the radial energy fluxes
provided by various physical processes, converted to luminosities and
normalized to the stellar luminosity.  The total luminosity $L(r)$ and its
components are defined by

\begin{equation}
F_e+F_k+F_{\rm r}+F_u+F_v+F_m=\frac{L(r)}{4\pi r^2},
\end{equation}
with
\begin{eqnarray}
F_e&=&\rb\, c_p\, \overline{v_r T'} \,,\\
F_k&=&\frac{1}{2}\, \rb\, \overline{v^2 v_r} \,,\\
F_r&=& -\kappa_{r}\, \rb\, c_p\, \frac{d\tb}{d r} \,, \\
F_u&=& -\kappa\, \rb\, \tb\, \frac{d\sb}{d r} \,, \\
F_v&=& -\overline{{\bf v}\cdot {\bf \cal D}} \,, \\
F_m&=& \frac{c}{4\pi}\, \overline{E_{\theta}B_{\phi}-E_{\phi}B_{\theta}} \,,
\end{eqnarray}
where the overbar denotes an average over spherical surfaces
and in time, ${\bf E}=4\pi\eta {\bf j} c^{-2} - ({\bf v}\times\BB) c^{-1}$ is the
electric field, $F_e$ the enthalpy flux, $F_k$ the kinetic energy
flux, $F_{\rm r}$ the radiative flux, $F_u$ the unresolved eddy flux,
$F_v$ the viscous flux and $F_m$ the Poynting flux.  The unresolved
eddy flux $F_u$ is the enthalpy (heat) flux due to subgrid-scale motions, which
in our LES-SGS approach takes the form of a thermal diffusion
operating on the mean entropy gradient.  The kinetic energy flux, the
viscous flux, the Poynting flux and the flux carried by unresolved motions are here all small
compared to $F_{e}$ and $F_{r}$.

The balance of energy transport is much as in our progenitor models in BBT.
As shown in Figure \ref{fluxbalance}, within case {\sl C4m} the enthalpy
flux is maximized near the middle of the convective core (at $r=0.08R$),
where it serves to carry about 50\% of the stellar luminosity, with the
remainder being transported by radiation.  Within the nearly adiabatic
stratification established in the convective core (with $\nabla -
\nabla_{ad} \sim 10^{-7}$), the associated temperature gradient serves to
specify a radiative flux $F_{r}$ that increases steadily with radius.  Thus
$F_e$ is forced to decrease in the outer half of the unstable core.  Beyond
the boundary of the convective core, the enthalpy flux becomes negative,
owing to the anti-correlation of radial velocity and temperature
fluctuations as penetrative motions are braked.  This inward directed
enthalpy flux is also manifested as a small dip in the total luminosity in
Figure \ref{fluxbalance}.  In real stars, or in fully relaxed simulations,
the radiative flux in that region would compensate for the negative
enthalpy flux.  However, our simulations have not been evolved for a
sufficiently long time to allow such adjustment to occur fully, since the
relevant thermal relaxation time is very much longer than other dynamical
timescales.  The small amplitude of the Poynting flux $F_m$ suggests that
although magnetic processes significantly impact the dynamics, they do not
actively transport enough energy to modify the radial energy flux balance
within the core.

\section{THE MANY SCALES OF FLOWS AND FIELDS}

The complex operation of the dynamo within the convective core generates
magnetic fields over a broad range of spatial scales, as evident in Figures
\ref{vrbrbp_Em} and \ref{vrbrbp_C4m}.  The manner in which the energy in
the fields and flows is distributed among these spatial scales, and between
axisymmetric and fluctuating components, provides perspectives on the
complicated nature of the magnetism.  Thus in addition to examining the
breakdown of these fields into their poloidal and toroidal components, we
also examine their spectral distributions and their probability density
functions.

\subsection{Mean and Fluctuating Magnetic Energy}

The strong magnetism generated in these simulations consists of both
mean (axisymmetric) and fluctuating fields.  We assess the balance
between these fluctuating and mean fields, defining various components of the magnetic energy as
\begin{eqnarray}
\mbox{MTE} &=&\frac{1}{8\pi}\left<B_{\phi}\right>^2 \,, \\
\mbox{MPE} &=&\frac{1}{8\pi}\left(\left<B_{r}\right>^2+\left<B_{\theta}\right>^2\right) \,, \\
\mbox{FTE} &=&\frac{1}{8\pi}\left((B_{\phi}-\left<B_{\phi}\right>)^2\right) \,, \\
\mbox{FPE} &=&\frac{1}{8\pi}\left((B_r-\left<B_{r}\right>)^2+(B_{\theta}-\left<B_{\theta}\right>)^2\right) \,, \\
\mbox{FME} &=&\frac{1}{8\pi}\left((B_r-\left<B_{r}\right>)^2+(B_{\theta}-\left<B_{\theta}\right>)^2+(B_{\phi}-\left<B_{\phi}\right>)^2\right) \,,
\end{eqnarray}
where we recall that the angle brackets $\left< ~ \right>$ denote a longitudinal average.  Here
MTE denotes the energy in the mean toroidal magnetic field, MPE likewise
that in the mean poloidal field, FTE the energy in the fluctuating toroidal
component, FPE that in the fluctuating poloidal field, and FME the total
energy in the fluctuating magnetic fields. In Figure \ref{magbalance}, we
illustrate for case {\sl Em} how these components (further averaged in
latitude) vary in strength with
radius throughout the convective 
core and the surrounding radiative envelope. The ME, TME, and PME are there 
averaged over a temporal interval of about 100 days representative of the 
extended plateau of high DRKE in Figure \ref{timetrace_Em}.  

The field within the core is mostly non-axisymmetric, with that fluctuating
field energy FME accounting for about 95\% of the total ME at most radii
within the convective core.  The remaining 5\% is distributed between the
toroidal and poloidal mean fields, with the former stronger there by about
a factor of two (Table 3).  The FME is in contrast divided in roughly equal
measure between FTE and FPE (not shown).  The balance of magnetic field
components changes rapidly near the edge of the radiative envelope (at
about $r=0.16R$).  Throughout the region of overshooting, the toroidal mean 
field becomes a steadily larger fraction of ME, whereas the fluctuating
field FME declines in proportional strength.  By $r=0.185R$, TME has become
as large as FME, and exterior to that radius it is the dominant contributor
to the magnetism.  The $m=0$ toroidal field energy MTE also remains much
larger than MPE through the region of overshooting and the radiative
envelope.

\subsection{Spectral Distributions of Flows and Magnetism}

The velocity and magnetic fields examined in Figures \ref{vrbrbp_Em} and
\ref{vrbrbp_C4m} for cases {\sl Em} and {\sl C4m} suggest that the magnetic
field posesses relatively more small-scale structure than the flows. This
is verified in Figure \ref{spectra} where we display for case {\sl Em} the
magnetic and kinetic energy spectra computed at two depths in the
convective core and within the region of overshooting.  The slope of the
magnetic energy spectrum (Fig. \ref{spectra}$b$) with degree $\ell$ is
much shallower than the kinetic energy spectrum (Fig. \ref{spectra}$a$) and
generally peaks at wavenumbers slightly higher. This means that the
magnetic energy equals or exceeds the kinetic energy at both intermediate
and small scales ($\ell \gtrsim 20$), even though when integrated over the
volume the magnetic energy is smaller than the kinetic energy.  Given that
our magnetic Prandtl number is greater than unity, such behavior is
expected in the range of wavenumbers located between the viscous and ohmic
dissipation scales, which here are in the range $\ell >100$.  More
surprising is that the magnetic energy also exceeds the kinetic energy over
a wide range of larger scales.  Possibly some guidance is afforded by
Grappin et al. (1983) in studying homogeneous isotropic MHD turbulent flows
in which there was overall equipartition between ME and KE.  They reported
that the difference between magnetic and kinetic energy spectra should
scale as $\ell^{-2}$ in the inertial range of the spectra, indicating a
dominant role of the magnetic field over the flow at small $\ell$'s. This
$\ell^{-2}$ scaling is not realized in our two cases here (except over a
small range of degrees in the overshooting layer), perhaps owing to the
effects of rotation and stratification not included in the Grappin et
al. (1983) analysis.  Throughout most of the convective core, both spectra
have broad plateaus at low wavenumbers, with shallow peaks near $\ell =
5$.  For degrees $\ell \gtrsim 10$, the spectra suggest some power law
behavior, but it extends only over a decade in degree so these simulations
do not possess an extended inertial range. The slope of the kinetic energy
spectrum (between $\ell^{-3}$ and $\ell^{-4}$) is substantially steeper
than that expected for homogeneous, isotropic, incompressible turbulence,
either with magnetic fields ($l^{-3/2}$) or without ($l^{-5/3}$) (e.g.,
Biskamp 1993).  The shallower magnetic energy spectra are somewhat closer
to the $\ell^{-3/2}$ behavior.

The energy contained in the dipolar and quadrupolar magnetic fields (i.e.,
modes $\ell=$1 and 2) is small when compared to the energy contained in all
the other modes.  In case {\sl Em} we find that they constitute about 5\%
of the total magnetic energy in the core, but contribute proportionately as
much as 15\% in the overshooting region. The quadrupolar field
is generally stronger than the dipolar one in the convective core by about a
factor of 2 to 3, but in the region of overshooting the dipole
term comes to dominate the quadrupole one.

In considering the energy spectra for KE and ME as a function of azimuthal wavenumber $m$
(not shown), within the convective core the dominant wavenumbers between 1
and 7 contain more power than the axisymmetric mode $m=0$, confirming the
predominantly non-axisymmetric nature of both the magnetic and velocity
fields.  Over the same temporal interval sampled by Fig. 15, the axisymmetric
$m=0$ represents about 3\% of the magnetic energy at $r=0.10 R$ and 5.5\% at
$r=0.05 R$.  We find similar but slightly smaller percentages in case {\sl
C4m}.  In the radiative zone, the axisymmetric mode becomes dominant for
the toroidal field and contributes about 29\% to the
magnetic energy contained in that layer. 

\subsection{Probability Density Functions}

The turbulent convective flows and magnetic fields in our simulations can
be further characterized by their probability density functions (pdfs).  In
idealized isotropic, homogeneous turbulence the velocity fields possess
Gaussian pdfs, yet departures from Gaussian statistics are known to be
present in many real turbulent flows.  In particular, velocity differences and
derivatives generally have non-Gaussian pdfs that are often
described by stretched exponentials $\exp[-\beta]$ with $0.5 \leq \beta
\leq 2$ (e.g., Castaing, Gagne \& Hopfinger 1990; Vincent \& Meneguzzi
1991).  The tails of the distributions are often nearly exponential ($\beta
\approx 1$) but can be even flatter, particularly in the viscous
dissipation range. Further, a flat slope ($\beta < 2$) indicates an excess of
high-amplitude events relative to a Gaussian distribution, a consequence of
spatial intermittency in the flow that may be associated with
the presence of coherent structures (e.g., Vincent \& Meneguzzi 1991;
Lamballais, Lesieur \& M\'etais 1997).

Figure \ref{pdfs_Em} shows pdfs for the radial and longitudinal components
of the velocity and magnetic fields for case {\sl Em} on a spherical
surface within the convective core ($r=0.10 R$) and the region of
overshooting ($r=0.16 R$).  The pdfs have been averaged over a 50 day
interval. In the convective core, the radial and longitudinal velocities
are nearly Gaussian with departure toward an exponential distribution in
their wings. By contrast, both components of the magnetic fields possess
strong departures from a Gaussian distribution, with pdfs closer to an
exponential distribution. In Figure \ref{pdfs_Em}d, the prominent hump in
the left wing of the $B_{\phi}$ distribution indicates that the toroidal
field is asymmetric and mostly negative over the temporal averaging
interval. In the overshooting region, the radial velocity $v_r$ (Fig.
\ref{pdfs_Em}e) is much less Gaussian than in the convective core, which
comes as a surprise since in BBT this was not the case. We find here that
the level of intermittency is higher than in our progenitor cases and that
the developed flows are much less steady due to the complex interaction
between convection and magnetic fields. This could perhaps partly justify
why $v_r$ is more intermittent at the convective core edge when magnetic
fields are present.  The longitudinal velocity $v_{\phi}$ is quite
asymmetric with a long tail for negative values, whereas the magnetic
fields are still non-Gaussian with a somewhat more intricate shape (less
smooth) than in the core, possibly revealing some long-living magnetic
structures in the overshooting layer. The pdfs within the core in case {\em
Em} (Figs. \ref{pdfs_Em}$a-d$) are qualitatively similar to those found
by Brun et al.  (2004) in the solar context, and by Brandenburg et
al. (1996) for compressible MHD convection in Cartesian geometries.

Higher order moments of the pdf, in particular the 3rd and 4th moments,
called respectively the skewness ${\cal S}$ and kurtosis (or flatness)
${\cal K}$ can be used to further quantify intermittency and asymmetry
(see Frisch 1995; Brun et al. 2004).  A large value for ${\cal S}$ indicates
asymmetry in the pdf whereas a large value of ${\cal K}$ indicates a high
degree of spatial intermittency.  For Gaussian pdfs, ${\cal S}=0$ and
${\cal K}=3$, whereas for exponential distributions ${\cal S}=0$ and ${\cal
  K}=6$. 

At $r=0.10 R$ the radial velocity is close to a Gaussian distribution with
(${\cal K}=3.6$) and possesses a relatively small negative skewness (${\cal
S}=-0.10$); the fastest downflows and upflows are of the same amplitude
$\sim$ 90 m s$^{-1}$ confirming the rather symmetric aspect of the
convective cells.  The longitudinal velocity $v_\phi$ is even more Gaussian
(${\cal K}=3.1$) but rather asymmetric (${\cal S}=0.75$), reflecting the
influence of the differential rotation. The radial and toroidal magnetic
fields are more intermittent than the velocity field (${\cal K}=$
5.9,11.8). The radial magnetic field $B_r$ appears to be quite symmetric
(${\cal S}=-0.10$), compared to $B_{\phi}$ that possesses a relatively
large skewness, ${\cal S}=-1.5$ mostly due to the presence of the
prominent hump in the left wing.  Maximum field strengths reach about
250 kG for the toroidal field and somewhat less (150 kG) for the radial
field.

At $r=0.16 R$ the radial velocity shows the greatest departures from a Gaussian, with ${\cal K}=19.5$, but is rather
symmetric (${\cal S}=0.09$).  The fastest downflows and upflows are of the
same amplitude $\sim$ 1 m s$^{-1}$ confirming the rather symmetric
aspect of the convective patterns in the overshooting region.  The zonal
velocity is still Gaussian (${\cal K}=3.6$) but even more asymmetric than
in the convective core (${\cal S}=-1.7$), reflecting the rather intricate
profile of differential rotation in that layer. The radial and toroidal
magnetic fields are somewhat less intermittent than in the core (${\cal
K}=$ 4.4, 3.7), confirming the stronger importance of the axisymmetric
part of the magnetic fields there. Both components are rather symmetric
with ${\cal S}=$-0.16, -0.09, respectively.  Maximum field strengths reach
about 45 kG for the toroidal field and much less (3 kG)  for the
radial field.

Case {\sl C4m} has pdfs, skewness and kurtosis values close to those for
case {\sl Em}. No clear trend due to a faster rotation rate is evident at
this stage.

\section{EVOLUTION OF GLOBAL-SCALE MAGNETIC FIELDS}

We now turn to considering the structure and evolution of the mean fields
realized in our simulations.  We here take these to be the $m=0$
(axisymmetric) component of the mostly non-axisymmetric magnetism generated
by dynamo action within the convective core.  We recognize that in seeking
to make contact with mean-field dynamo theories, other spatial and temporal
averaging could be employed, such as general averaging over intermediate
scales.  However defined, such large-scale fields have particular
significance in stellar dynamo theory.  Our results provide insight into
the generation of mean magnetic fields by turbulent core convection and
might be used to evaluate and improve mean-field dynamo models that do not
explicitly consider the turbulent field and flow components (e.g.\ Krause
\& R\"adler 1980, Moss 1992, Ossendrijver 2003). We define the mean
poloidal magnetic field to be the longitudinally-averaged radial and
latitudinal components, $\left<{\bf B_p}\right> = \left<B_r\right> \uvr +
\left<B_\theta\right> \uvt$, and the mean toroidal field in terms of the
longitudinal component $\left<B_t\right>\uvp = \left<B_\phi\right> \uvp$.

The mean toroidal fields in our simulations can arise from the shearing,
stretching, and twisting of mean and fluctuating poloidal fields by
differential rotation (the $\omega$-effect), or from helical convective
motions (the $\alpha$-effect).  In contrast, mean poloidal fields are
generated from fluctuating toroidal fields only via the $\alpha$-effect.
Thus the mean and fluctuating magnetic fields, the differential rotation,
and the convective flows are intimately linked.

\subsection{Axisymmetric Poloidal Fields}

The energy contained in the axisymmetric poloidal field throughout the shell is of
order 5\% of ME in the core and much less ($<0.1\%$) in the radiative envelope.
Typical poloidal field stengths are respectively of order 300 G and 0.1 to 1
G.

Figure \ref{Bpol} illustrates the structure and evolution of the axisymmetric
poloidal field in case {\sl Em}.  The top row shows four snapshots of the
magnetic lines of force of $\left<{\bf B_p}\right>$ within the convective
and radiative domains.  Such a mean poloidal field within the core shows
intricate morphology, with islands of positive and negative polarity that
often intermix.  During some intervals the field is dominated by a single
polarity (Fig. \ref{Bpol}$a$, $d$), whereas at other times both polarities
are present in roughly equal measure (Fig. \ref{Bpol}$b$,$c$).  This
complex evolution is connected to the non-axisymmetric nature of the
convective flows that have given rise to these fields from their initial
weak dipole state.  The evolution of $\left<{\bf B_p}\right>$ in the
radiative envelope is more passive, and depends strongly on the properties
of that field at the interface with the convective core and their ability
to diffuse outward, as evinced by $\left<{\bf B_p}\right>$ now differing from
its initial dipolar configuration. Within the core, the presence of strong magnetic field
gradients and magnetic diffusion lead to continous reconnection of the
magnetic field lines. Such reconnection can be seen in the sequence within
Figures \ref{Bpol}\emph{b,c,d} where in the northern hemisphere (at low
latitudes near the core boundary) reconnection between fields of differing
polarities occurs, resulting in a small isolated loop of positive polarity
at mid latitudes that later rises slowly and diffuses away.

The regeneration of magnetic flux by the convection can lead to global
reversals of the magnetic field polarity, as seen in Figures
\ref{Bpol}$a-d$. Figure \ref{Bpol}$e$ shows the temporal evolution of the
average polarity of the poloidal field in case {\sl Em}, which we define in
terms of the radial magnetic field $B_r$ averaged over the northern
hemisphere both at the convective core boundary (denoted by the solid line)
and at the top boundary (dashed line). Figure \ref{Bpol}$f$ is the
equivalent plot for case {\sl C4m}. This measures the total magnetic flux
that passes through the northern hemisphere at those radial surfaces.
Positive values indicate that the field is outward on average in the
northern hemisphere, as in the dipolar seed field. 

Figure \ref{Bpol}$e$ shows the evolution of the average field polarity in
case {\sl Em} between 4500 and 7000 days of computed physical time,
corresponding to an interval in which the magnetic energy has reached a
statistically stable phase.  Two field reversals occur on a time scale of
about 1000 days, but we cannot assess whether such polarity reversals are
likely to be continued and regular. In the radiative envelope reversals
could occur, but on a much slower time scale as fields diffuse upwards. The
behavior in Figure \ref{Bpol}$f$ for case {\sl C4m} is similar to that seen
in case {\sl Em}. Such changes in the magnetic polarity in the convective
domain have also been seen in Brun et al. (2004) in the solar
context. There also the convection generates rather weak axisymmetric fields and
the fluctuating fields are the dominant players.

\subsection{Axisymmetric Toroidal Fields} 

The axisymmetric toroidal field in the convective core contains about 6 to 10\% of
the total magnetic energy, about a factor of two larger than the energy in
the axisymmetric poloidal field.

Figure \ref{Btor} shows two snapshots of the radial and latitudinal
variation of the longitudinally-averaged toroidal magnetic field $\left<B_t\right>$ for case
{\sl Em} at times coincident with Figures \ref{Bpol}\emph{a},\emph{d}.  We
can see that $\left<B_t\right>$ possesses small-scale structure, with
little correspondence apparent between the two time samples, indicating the
complex evolution of the axisymmetric toroidal field.  Mixed polarities and intricate
topologies are present throughout the convective core.  Varying symmetries
may be evident at different instants, but do not persist over extended intervals.

Some hints regarding the interplay between the $\alpha$-effect and the
$\omega$-effect in generating axisymmetric toroidal fields are afforded by
comparing Figure \ref{Btor} with the views of $\left<{\bf B_p}\right>$ in Figure
\ref{Bpol}.  If the $\omega$-effect had a dominant role in the generation
of $\left<B_t\right>$, as may be realized at the core boundary where
convective motions have waned, the evolution of the axisymmetric poloidal and
toroidal fields would be clearly linked.  The largely retrograde
differential rotation acting on a negative poloidal field structure would
generate a toroidal field with two opposite polarities: negative in the
lower part of the structure and positive in the upper part.  That the
non-axisymmetric convection also plays a role in generating
$\left<B_t\right>$ through the $\alpha$-effect obscures the connection
between structures in the two fields, yet some links are indeed
revealed by Figure \ref{Btor}. In Figure \ref{Bpol}$a$ the poloidal field
possesses a counterclockwise (negative) polarity at mid latitudes along the
convective core boundary in the northern hemisphere.  The appearance in
Figure \ref{Btor}$a$ of both senses of toroidal fields at the same location
may be indicative of the $\omega$-effect at work.  Similar linkages are
apparent in Figure \ref{Btor}$b$, where the corresponding poloidal field
was largely of negative polarity along the core interface but
$\left<B_t\right>$ is largely of differing senses in the northern and
southern hemispheres.  Of course some time lag should exist between the
establishment of toroidal mean fields from a given mean poloidal field
configuration via the $\omega$-effect, further complicating the
interpretation of links between the two fields. Furthermore, many
departures from this idealized description of the generation of fields
occur due to the major role played by the $\alpha$-effect within the core
in giving rise to the magnetism.

\subsection{Wandering of the Poles}

Like the axisymmetric poloidal and toroidal fields, the dipole component of the
magnetism attracts interest despite its relatively small amplitude.  Dipole
magnetic fields have figured prominently in some theoretical efforts to
construct simplified models of the interiors of A-type stars (e.g., Mestel
\& Moss 1977).  In addition, the presence at the surface of largely
dipolar fields further serves to motivate the examination of such fields
in our simulations, though this interior magnetism may well be screened
from view by the extensive radiative envelope.  We here assess the temporal
variations of the dipole field, which may differ appreciably from those of
the rapidly evolving and intricate small-scale fields.

In our two simulations, the maximum amplitude of the dipole magnetic field
(namely the $\ell=1$ component of $B_r$) is generally no greater than about
5\% of the maximum total radial field, with variations in strength by a
factor of two occurring as the fields evolve in time.  Further, the low spherical
harmonic degrees $\ell = 2-12$ typically possess somewhat greater amplitudes
than the dipole. The evolution in case {\sl C4m} of the dipole axis over a
period of $\sim$1900 days (or about 270 rotations) is assessed in Figure
\ref{polewander}, showing the position in latitude and longitude of the
positive dipole axis with time.  The orientation of the dipole, which is
inclined with respect to the rotation axis, varies slowly: during the
lengthy interval sampled, the pole completes only two full revolutions
around the rotation axis, though there are brief periods during which its
movement is more rapid.  The wanderings from the northern hemisphere into
the southern and back are similarly leisurely: only three such inversions
of polarity, each separated by more than 500 days, are visible in Figure
\ref{polewander}.  The orientation of the dipole appears to correspond very
well with the sign of the axisymmetric poloidal magnetic field when integrated
over the northern hemisphere (Fig. \ref{Bpol}$f$).
 As the dipole meanders from a northerly orientation to a southerly one,
the sign of the integrated radial field at the edge of the convective core
also flips from positive to negative.  Similar slow wanderings are observed
in case {\sl Em}, though there the dipole axis lies close to the equatorial
plane over the first 1500 days of the interval sampled in Figure
\ref{Bpol}$e$.  Thus the slow evolution of the dipole component of the
magnetism stands in contrast to the far more rapid changes seen in the
high-degree components.  

\section{SOME ASPECTS OF FIELD GENERATION}

The detailed manner in which sustained dynamo action is achieved in our
models is challenging to understand, since we have relatively few theoretical
tools for predicting such behavior short of carrying out nonlinear
simulations.  In mean-field dynamo theory (see e.g., Moffatt 1978; Brandenburg \&
Subramanian 2004), one commonly speaks of the $\alpha$-effect, by which
helical turbulence in a resistive medium can produce mean toroidal magnetic
fields from seed poloidal ones and vice versa, and of the $\omega$-effect,
in which stretching of field lines by contrasts in angular velocity can
generate mean toroidal fields from poloidal ones.  Although the $\omega$
and $\alpha$ effects strictly refer only to the generation of mean toroidal
and poloidal fields from mean and fluctuating fields, their counterparts in
the equation for the evolution of the fluctuating fields may be useful in looking at the
generation of the strong fluctuating fields realized in our simulations.

Among these generation terms, the G-current ${\bf G}={\bf v}' \times {\bf
B}' - <{\bf v}' \times {\bf B}'>$ plays a pivotal role. In the traditional
first-order smoothing approximation of mean field dynamo theory (Krause \&
R\"adler 1980, Ossendrijver 2003), this term is neglected, providing a
simple closure procedure for the mean field induction equations. We find in
our simulations that the G-current is by no means small, with $<{\bf v}'
\times {\bf B}'>$ considerably smaller (about only 5\%) than $ {\bf v}' \times {\bf B}'$.
Further, the nonlinear dynamo action realized in our simulations induces
preferentially strong non-axisymmetric fluctuating magnetic fields rather
than axisymmetric ones, with the latter having only a very weak dynamical
role. Our core convection dynamo simulations suggest that higher-order mean
field dynamo theories, which do not neglect the G-current, may be required
to explain the dynamo operating in a stellar convective core.

A physical quantity of some interest in analyzing the properties of the
magnetic field generated in our simulations is the kinetic helicity ${\bf
v}\cdot(\nab\times{\bf v})$.  In Figure 20, we display the kinetic helicity
both for case {\sl Em} and for its hydrodynamical progenitor as a function
of radius, averaged over the northern hemisphere and in time.  It is
negative in most of the domain except for a positive region near the center
of the core.  In contrast, the current helicity ${\bf j}\cdot{\bf B}$ in
our simulations shows no comparable trends or sign preference in a given
hemisphere, in agreement with the results obtained in the solar dynamo
simulations of Brun et al. (2004).  As a consequence, we see no evident
relation between the kinetic and magnetic helicities in our modeling,
though some links are implicit in certain mean field theories (e.g.,
Ossendrijver 2003).  Turning back to Figure 20, we note that the kinetic
helicity in case {\sl Em} possesses a smaller amplitude than in its
hydrodynamic progenitor.  This suggests that the magnetic field acts to
reduce local shear and stretching, in particular near sites of strong
vorticity, leading to a reduced helicity in the convective region. Outside
the core, in the region of overshooting and beyond, the kinetic helicity is
very small, since only weak fluid motions persist.  Thus the generation of
magnetic fields by helical convective motions must also basically vanish
outside the core.

Conversely, angular velocity contrasts that are weak within the convective
core grow stronger in the region of overshooting, particularly in case {\sl
Em} with its interface of strong shear. Thus the balance of toroidal and
poloidal field should vary with radius, as the relative importance of the
helical motions grows smaller and the contribution of the large-scale shear
becomes larger.  The radial variation of the energy in the fluctuating and
axisymmetric magnetic fields (\S5.1) therefore provides clues
about the mechanisms responsible for building the magnetism. Although the
total ME declines sharply outside the convective core, the axisymmetric
toroidal field $\hat{B}_{\phi}$ within the region of overshooting is still
considerable.  This, together with the longitudinally elongated topology of
$\hat{B}_{\phi}$ (Fig. \ref{hvr}$c$), indicates that the large-scale shear
helps generate the magnetism.  That TME exceeds PME by a factor of two
within the convective core suggests that the equivalent of an
$\omega$-effect plays a role there as well.  These core convection dynamos
thus generate magnetic fields through the joint effects of large-scale
shear and helical motions acting on the axisymmetric and the
non-axisymetric fields. The large-scale shear appears to dominate the
generation of field near the convective core boundary, while the helical
motions generate fields in a more distributed manner within the core.


\section{CONCLUSIONS AND PERSPECTIVES}

The simulations here reveal that vigorous convection within the cores of
rotating A-type stars can serve to build strong magnetic fields through
dynamo action.  Small initial seed magnetic fields are amplified in
strength by many orders of magnitude and sustained against ohmic decay,
ultimately yielding fields that are nearly in equipartition with the flows.
The resulting highly time-dependent magnetism possesses structure on many
scales, with $B_r$ mainly fibril and $B_{\phi}$ stretched by the zonal
flows (differential rotation) into large-scale bands that extend around the
core.  Within the core, the magnetism is predominantly fluctuating, with
such non-axisymmetric fields accounting for about 90\% of the total magnetic
energy.  The accompanying weak mean (axisymmetric) fields evolve comparatively slowly,
undergoing flips in average polarity on time scales of hundreds of days.  In
the farther region of overshooting and beyond, where the magnetic energy
plummets from its interior value, the mean toroidal field becomes the
dominant component of the surviving magnetism.

\begin{figure}[hpt]
\center
\epsscale{0.64}
\plotone{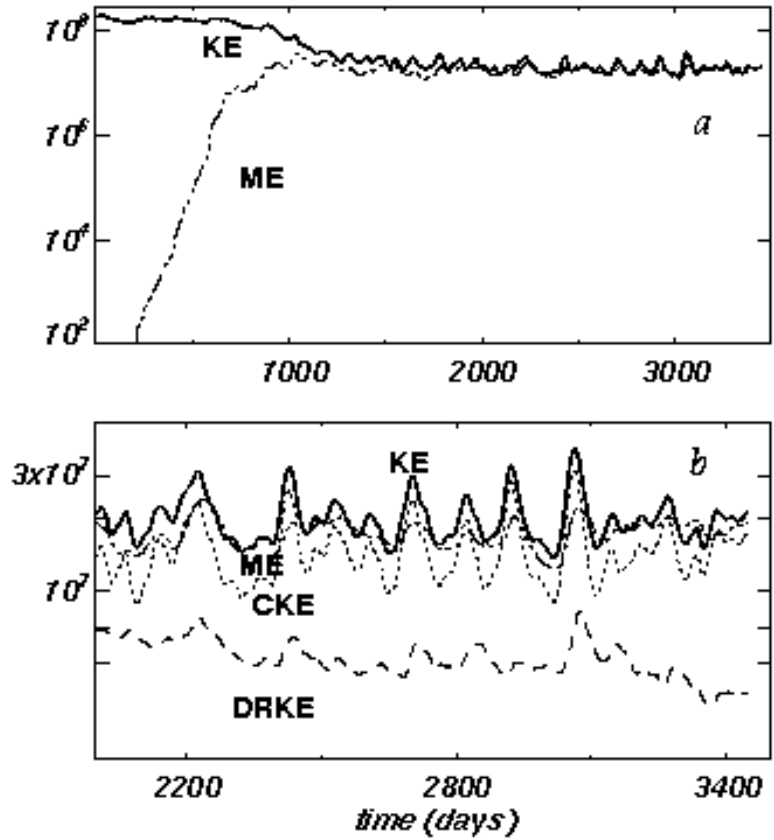}
\caption{\label{timetrace} Temporal evolution in case {\sl C4m} of the
  volume-averaged total kinetic energy density (KE) and the magnetic energy density (ME).
  ($a$) The initial seed magnetic field is amplified by many orders of
  magnitude.  After an initial phase in which ME grows exponentially, it
  equilibrates to a level in which it becomes comparable to KE, which has
  been lessened by the feedback of the magnetism upon the flows.  ($b$)
  Detailed view of fluctuations of energy densities once equilibration is
  approached, showing also energy densities of the convection (CKE)
  and the differential rotation (DRKE).}
\end{figure}

\begin{figure}[hpt]
\center
\epsscale{1.0}
\plotone{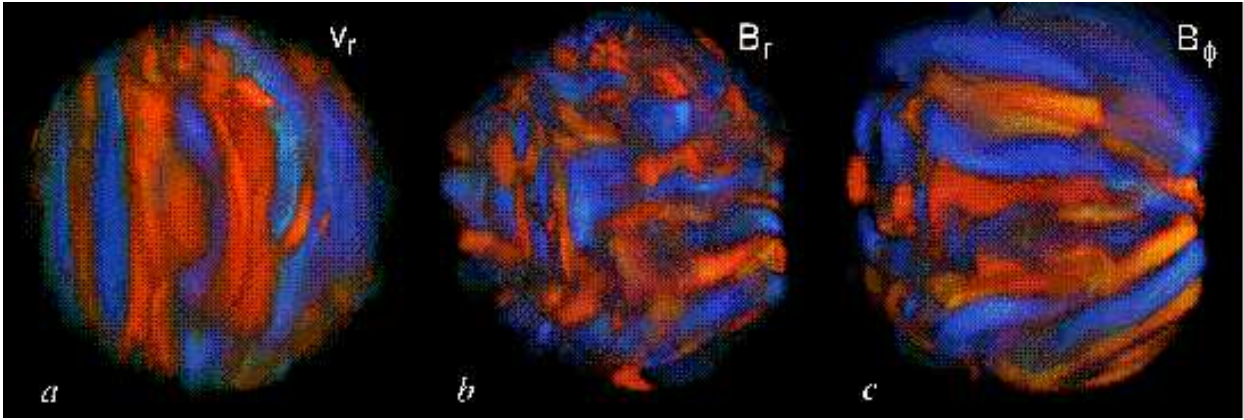}
\caption{\label{hvr} Volume renderings of flow and magnetic structures at
  one instant in case {\sl C4m} near the outer boundary of the prolate
  convective core.  ($a$) Radial velocity $v_r$ exhibits columnar
  structures aligned with the rotation axis (here oriented vertically).
  Little asymmetry is apparent between upflows (reddish) and downflows
  (bluish).  ($b$)  The radial magnetic field $B_r$ is more tangled, with
  field polarity shown in contrasting tones.  ($c$)  The longitudinal magnetic
  field $B_{\phi}$ possesses a distinctive ribbon-like morphology, with
  coherent bands that extend around much of the core.}
\end{figure}

\begin{figure}[hpt]
\center
\includegraphics[width=16.5cm]{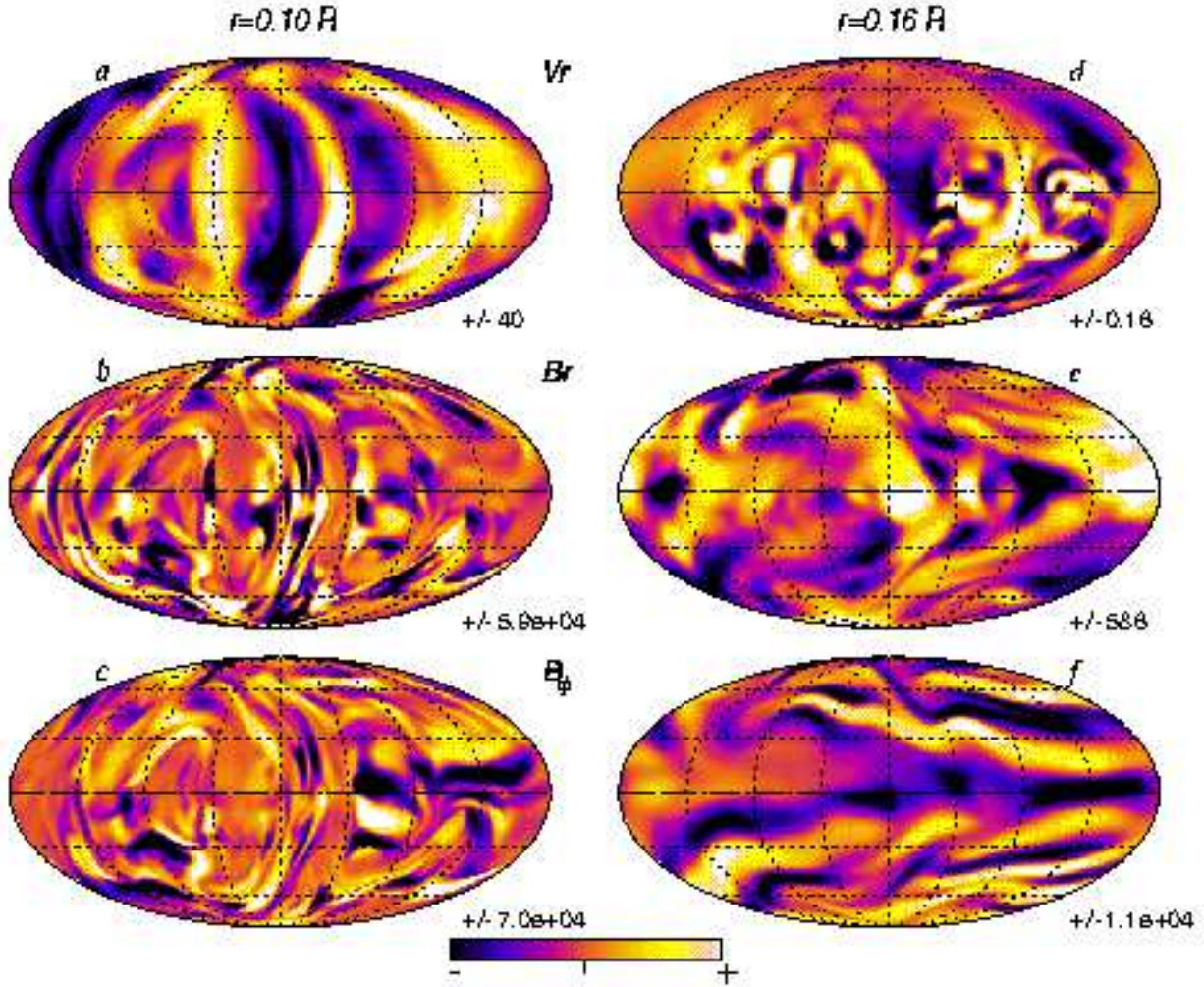}
\caption{\label{vrbrbp_Em} Global mappings at one instant in case {\sl Em}
  of $v_r$, $B_r$, and $B_{\phi}$ sampled on two spherical surfaces, both
  at mid-core ($r=0.10 R$, \emph{left}) and in region of penetration and
  overshooting ($r=0.16 R$, \emph{right}).  Shown are Mollweide projections
  with the dashed horizontal line denoting the equator.  All fields share
  the same symmetric color table, with positive values in bright tones and
  negative ones in dark tones.  The amplitude ranges are
  indicated adjacent to each panel, with magnetic fields in G and
  velocities in m s$^{-1}$.}
\end{figure}

\begin{figure}[hpt]
\center
\epsscale{1.0}
\includegraphics[width=16.5cm]{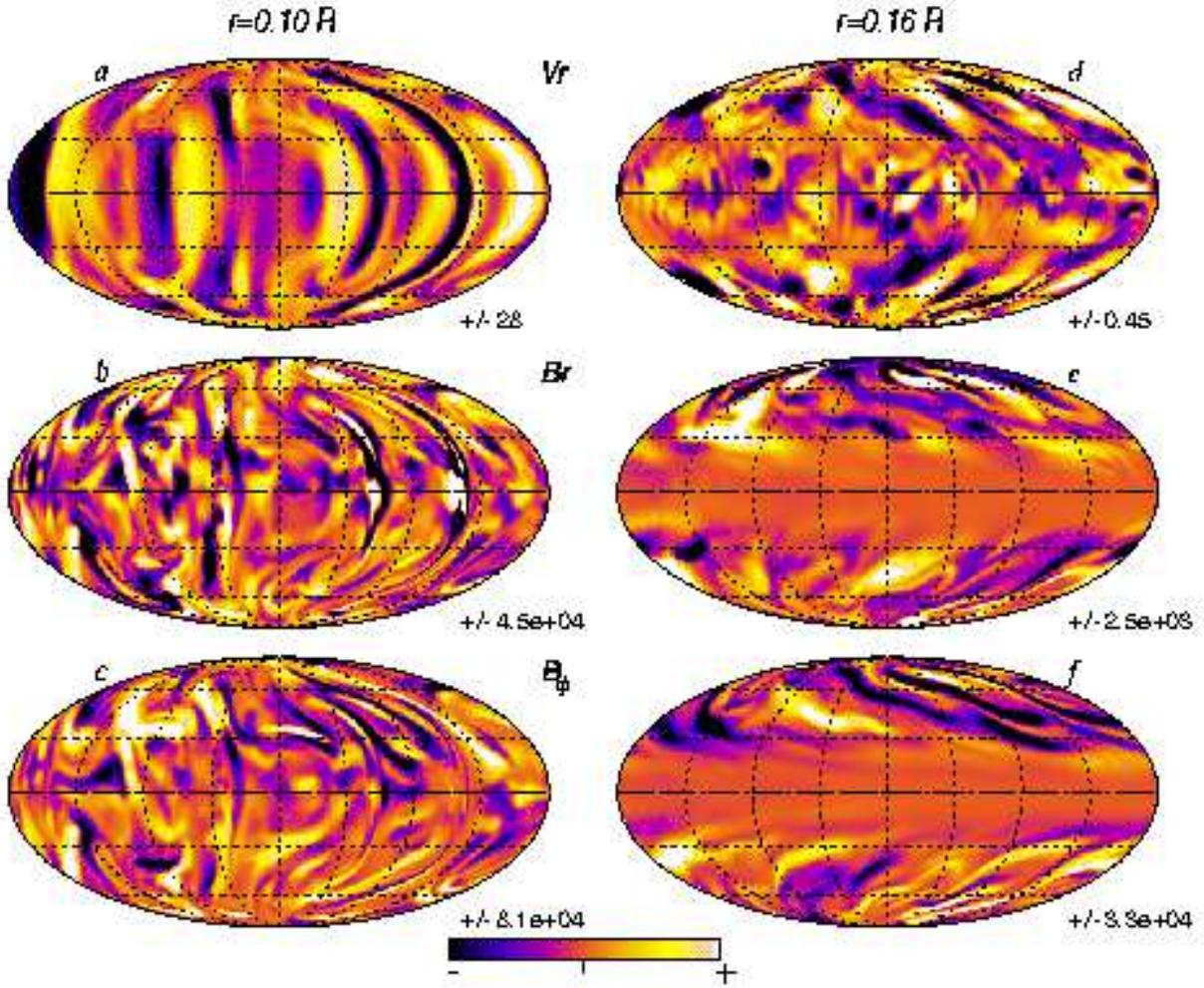}
\caption{\label{vrbrbp_C4m} As in Fig. \ref{vrbrbp_Em}, global Mollweide
  projections of $v_r$, $B_r$,
  and $B_{\phi}$ for the more rapidly rotating case {\sl C4m} at one instant in time.}
\end{figure}

\begin{figure}
\center
\includegraphics[width=2.5in, trim= 72 0 62 0]{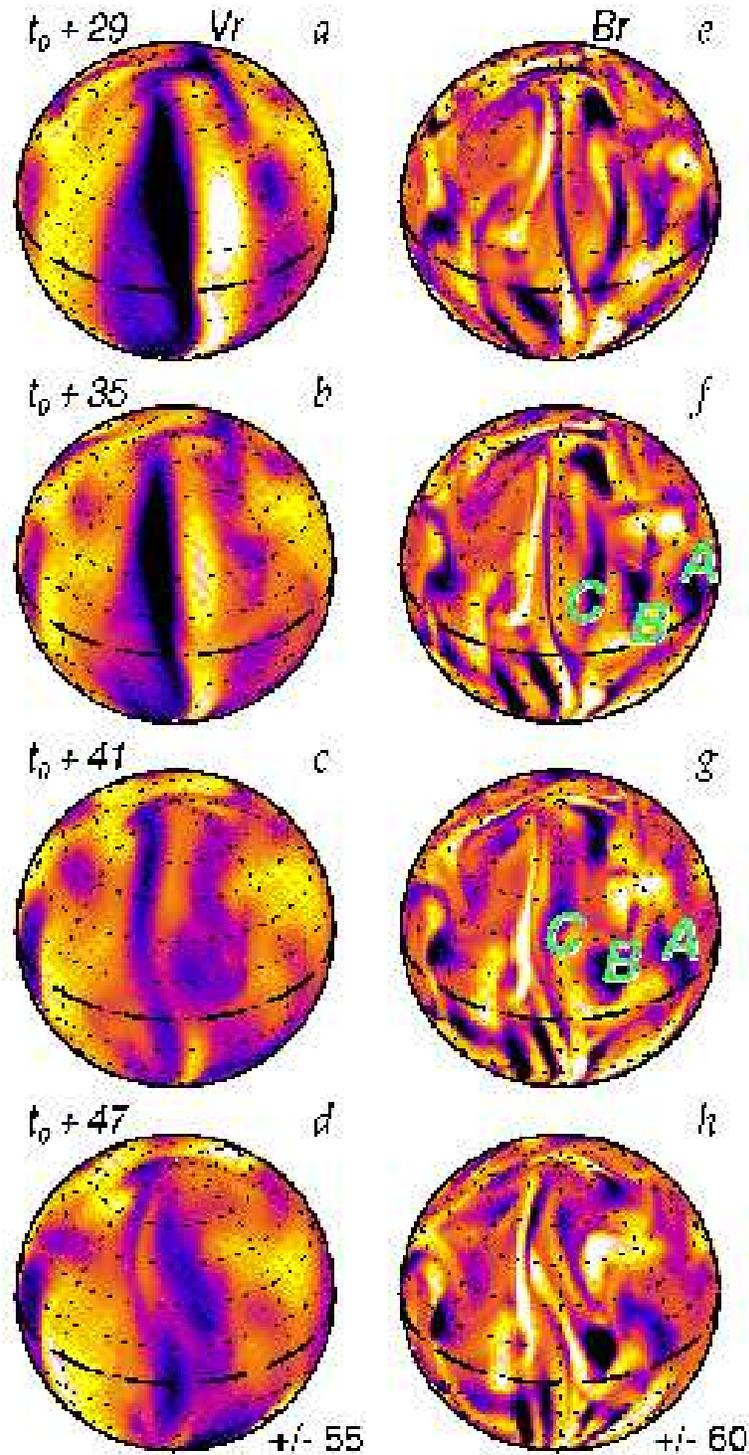}
\caption{\label{timevol_Em} Rapid sampling in time of the evolution of
  structures seen in $v_r$ (\emph{left}) and $B_r$ (\emph{right}), as
  viewed on spherical surfaces at mid-core ($r=0.10 R$) for case {\sl Em}.
  The four snapshots are separated by 6 days each, starting from a mature
  time $t_o$ within the simulation.  Features ({\sl A}, {\sl B}, {\sl C},
  \emph{labeled}) in the flows and magnetism persist, but are advected and
  sheared, propagate relative to the frame, and can cleave into smaller
  structures.  The color table is as in Fig. \ref{vrbrbp_Em}, and scaling
  is indicated.}
\end{figure}

\begin{figure}[hpt]
\center
\epsscale{0.93}
\plotone{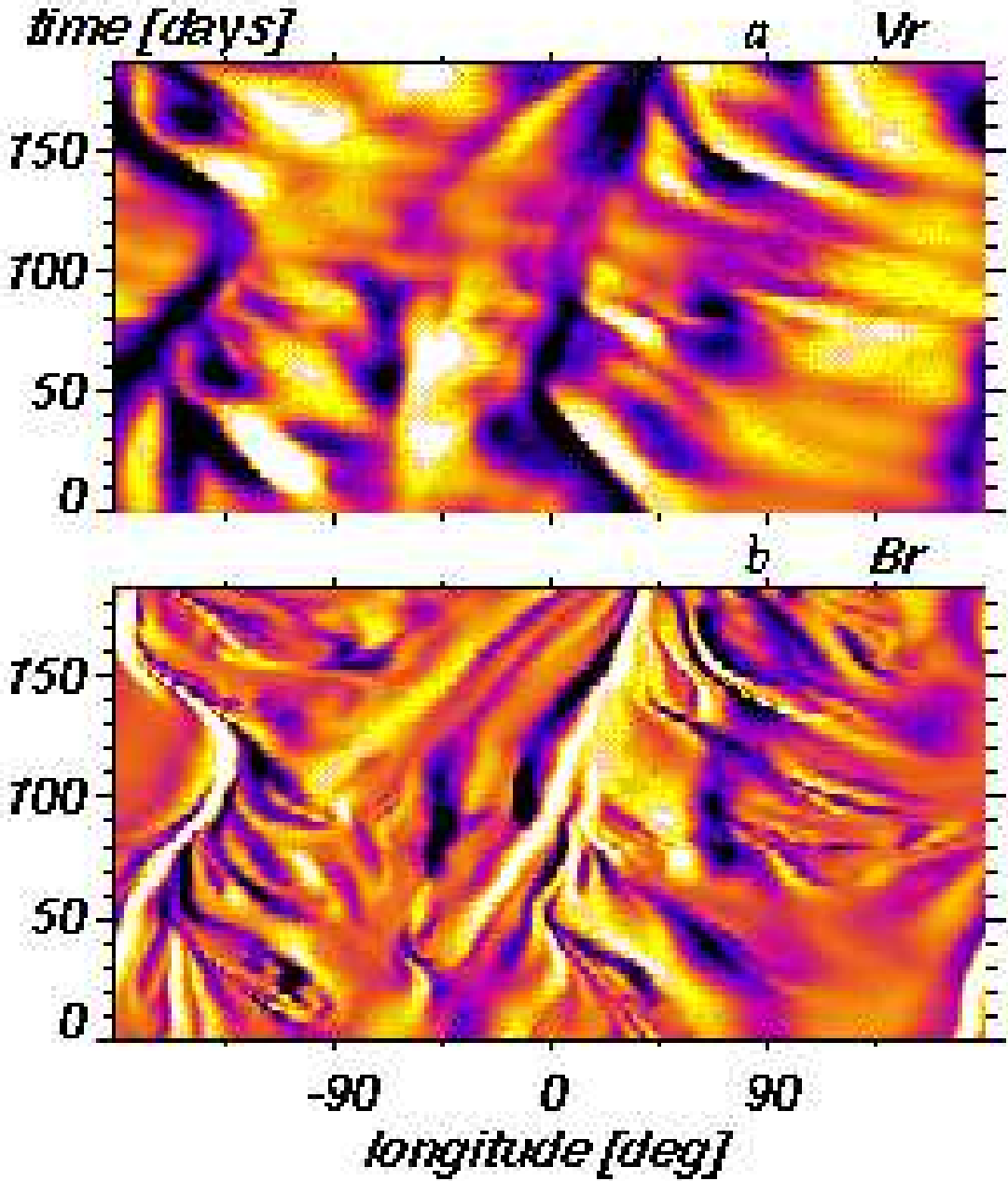}
\caption{\label{timelongEm_eq} Extended evolution and propagation assessed
  by time-longitude sampling of ($a$) $v_r$ and ($b$) $B_r$ at mid-core
  ($r=0.10 R$) in case {\sl Em} at the equator.  At such low latitudes,
  persistent features in both $v_r$ and $B_r$ tend to propagate prograde
  (to the right) in longitude (relative to the frame).  There is close
  correspondence in the structures evident in $v_r$ and $B_r$.  The color
  table and scaling is shared with Fig. \ref{timevol_Em}, as is the
  sampling starting time $t_o$.}
\end{figure}

\begin{figure}[hpt]
\center
\epsscale{0.93}
\plotone{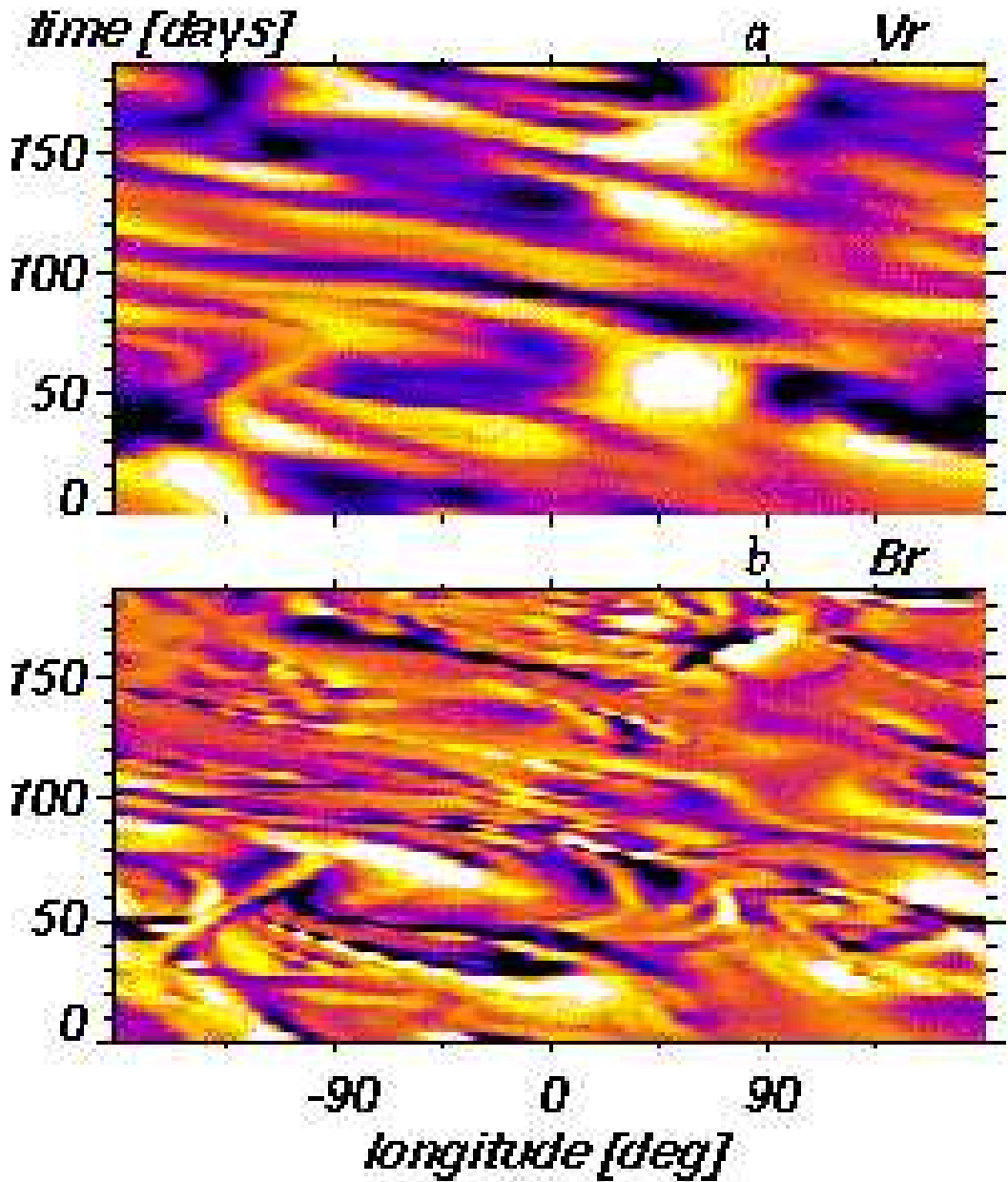}
\caption{\label{timelongEm_60} As in Fig. \ref{timelongEm_eq}, but
  time-longitude maps for $v_r$ and $B_r$ at latitude 60\degr. Here the propagation of features in both fields is distinctly
retrograde relative to the frame.} 
\end{figure}

\begin{figure}[hpt]
\center
\epsscale{0.60}
\plotone{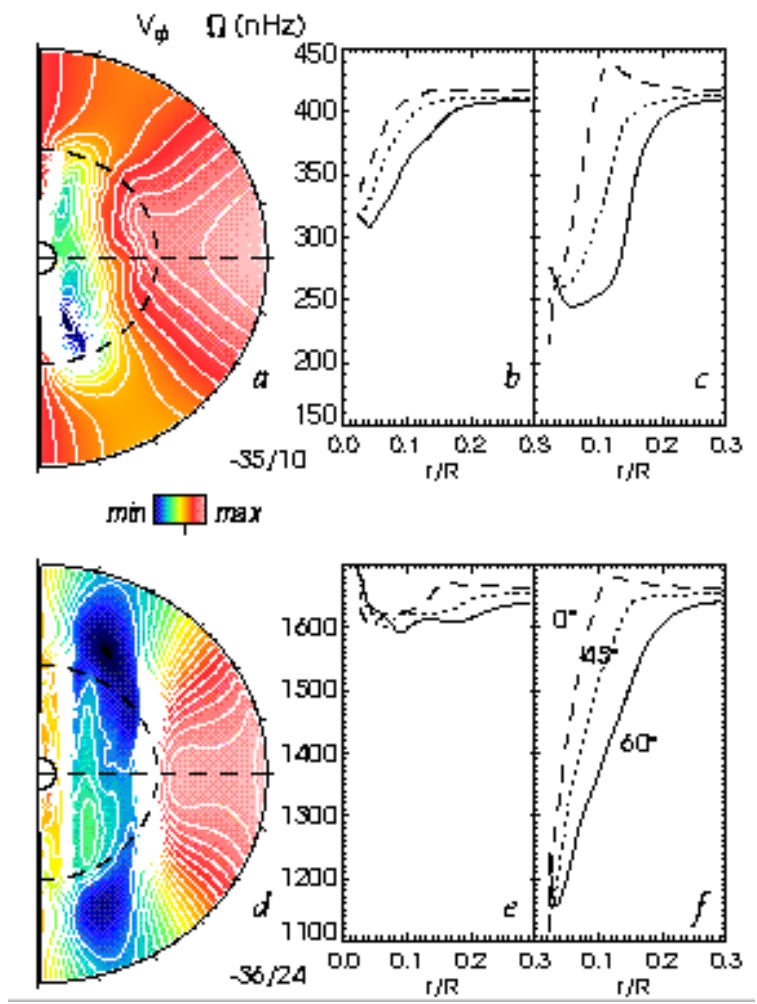}
\caption{\label{diffrotn_maghydro} Differential rotation established in
case {\sl Em} (\emph{top}) and {\sl C4m} (\emph{bottom}), and their progenitors.  ($a$,$d$)
Mean zonal velocity $\vph$, averaged in time and longitude, as contour plots in
radius and latitude, with color bar and ranges (in m s$^{-1}$) indicated.
Equator is horizontal, rotation axis is vertical, and outer extent of
convective core is indicated by dashed curve.  ($b$,$e$) Angular velocity
$\hat{\Omega}$ with radius for latitudinal cuts at 0\degr,45\degr, and
60\degr.  ($c$,$f$) $\hat{\Omega}$ achieved in progenitor nonmagnetic
models.  The magnetism acts to inhibit the strong angular velocity contrasts
realized in the progenitor simulations.}
\end{figure}

\begin{figure}[hpt]
\center
\epsscale{0.64}
\plotone{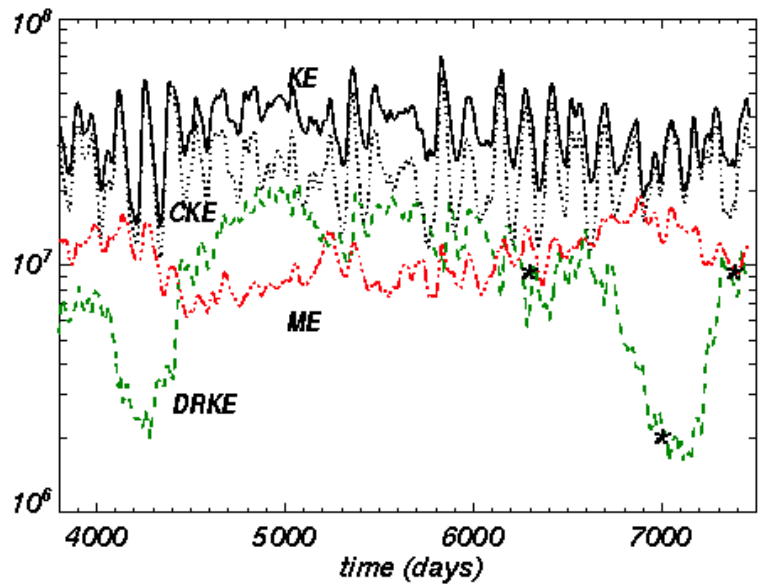}
\caption{\label{timetrace_Em} Detailed view in case {\sl Em} of variations
in the volume-averaged energy densities of the convection (CKE),
differential rotation (DRKE), total kinetic energy density (KE), and
magnetic energy (ME).  Here DRKE undergoes two pronounced minima, the
beginnings of which coincide with times at which ME climbs above $\sim 1.2
\times 10^7$ erg cm$^{-3}$, or about 40\% of KE.  Indicated on the DRKE
trace are the three times sampled for the differential rotation snapshots
in Fig. \ref{3omegas}.}
\end{figure}

\begin{figure}[hpt]
\center
\epsscale{0.60}
\plotone{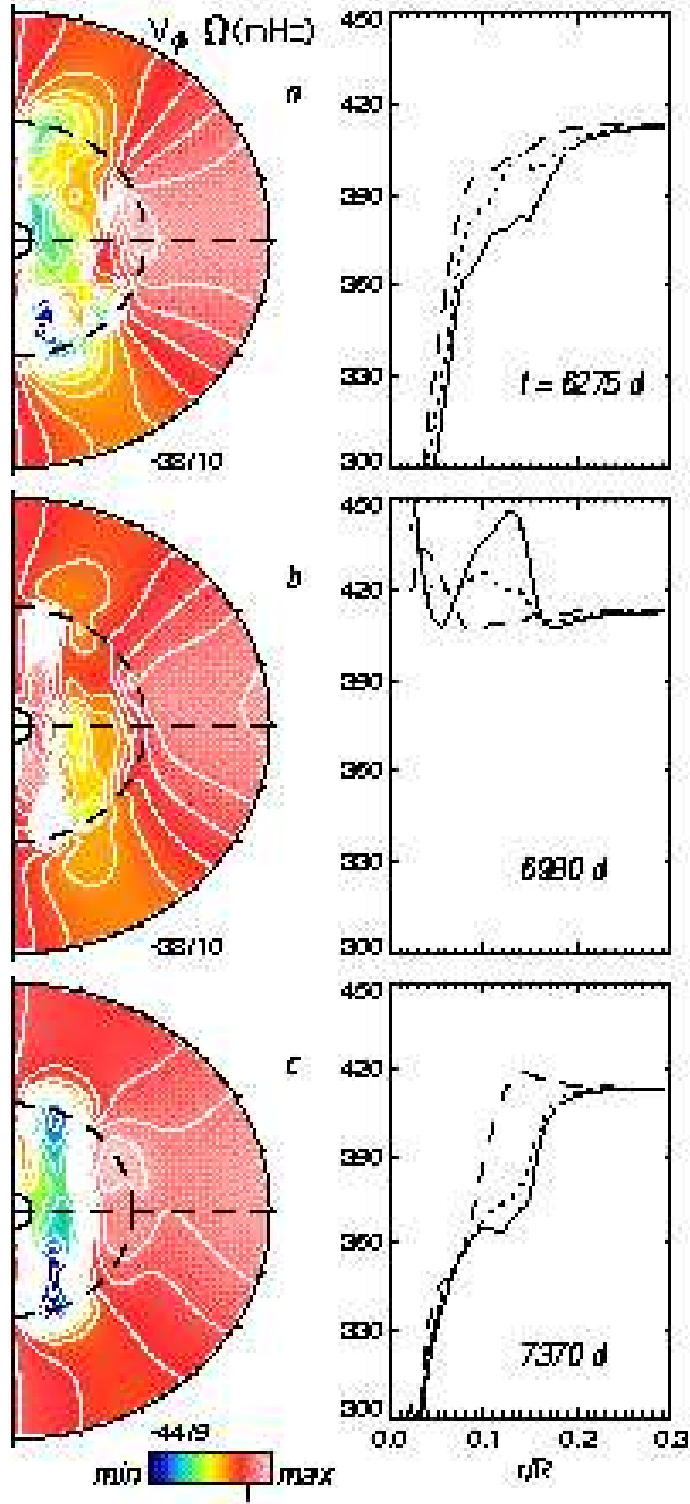}
\caption{\label{3omegas} Differential rotation achieved in case Em at three
different time sampling intervals before ($a$, upper), during ($b$,
middle), and after ($c$, lower) a grand minimum in DRKE in
Fig. \ref{timetrace_Em}.  Shown as contour plots (left) are the longitudinal averaged
zonal velocity $\vph$ averaged over brief (20 day) intervals, accompanied by (\emph{right})
angular velocity $\hat{\Omega}$ as radial cuts at the three latitudes indicated.}
\end{figure}

\begin{figure}[hpt]
\center
\epsscale{1.0}
\plotone{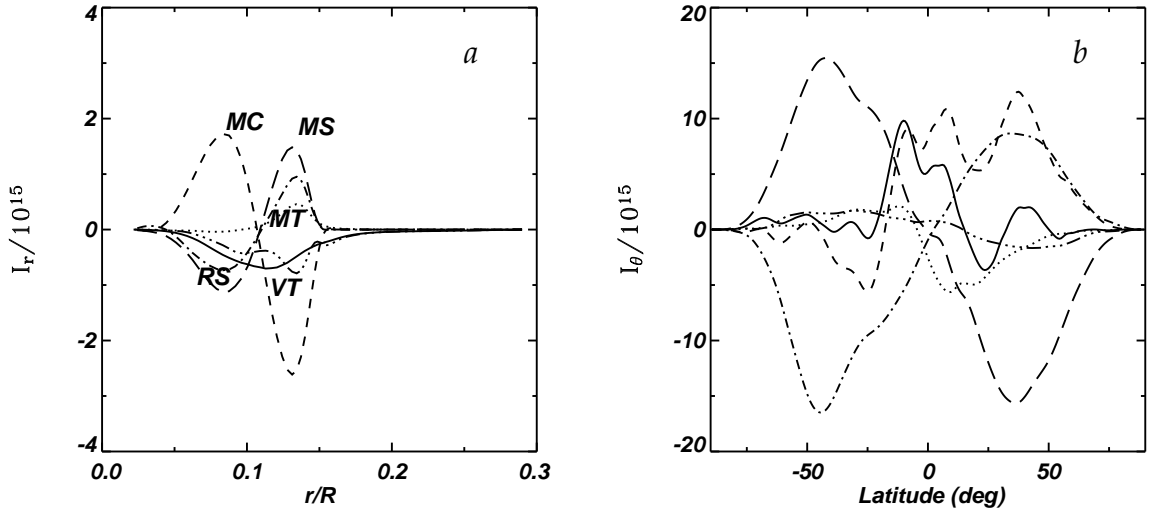}
\caption{\label{amom_balance} Temporal average in case {\sl Em} of
(\emph{a}) the integrated vertical angular momentum flux $I_r$ and
(\emph{b}) the integrated latitudinal angular momentum flux $I_{\theta}$.  These
have been decomposed into components due to viscous transport (labeled VT), Reynolds stress (RS),
meridional circulation (MC), Maxwell stress (MS), and large-scale magnetic
torque (MT), and the solid curves represent
the total fluxes.  Positive quantities represent fluxes radially outward,
or latitudinally from north to south.  The interval chosen for the time
averages spans 300 days late in the simulation.}
\end{figure}

\begin{figure}[hpt]
\center
\epsscale{0.60}
\plotone{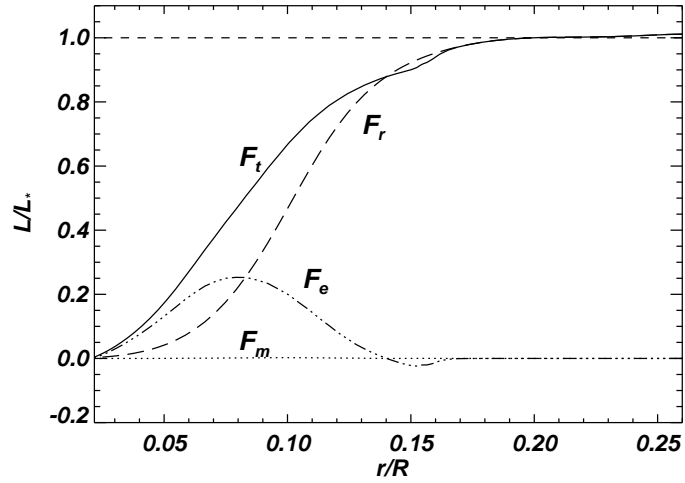}
\caption{\label{fluxbalance} Variation with radius of the radial transport
of energy in case {\sl C4m}, as averaged over an interval of about 60 days.
Shown are the enthalpy flux $F_{e}$, the radiative flux $F_{r}$, and the
Poynting flux $F_{m}$, together with the total flux $F_{t}$; all quantities
have been expressed as luminosities. The convective core extends here to
about $r=0.14R$, with the positive $F_{e}$ there serving to carry as much
as 80\% of the total flux.  The further region of overshooting involves a
small negative (inward directed) enthalpy flux.  Here $F_{m}$ is small
throughout the domain.}
\end{figure}

The differential rotation established in the hydrodynamic progenitors is
lessened in latitude by the presence of magnetism, as the strong Maxwell
stresses associated with the fluctuating fields transport angular momentum
poleward, with the large-scale magnetic torques playing only a small role in the
overall latitudinal balance.  Conversely, the radial transport of angular
momentum by the Maxwell stresses does not oppose that of the Reynolds
stresses throughout much of the convective core.  Thus the Maxwell
stresses are found to play a significant role in the angular momentum
transport.  In case {\sl Em}, with ME about 40\% of KE, central columns of
slow rotation are still realized, with considerable variations in strength
as the simulations evolve.  Oscillations seen in the energy densities DRKE
and ME (Fig. \ref{timetrace_Em}) hint at the intimate connection between
magnetic fields and flows, with intervals of high ME apparently acting to
quench the differential rotation; the resulting weak angular velocity contrasts
eventually lead to decreases in ME.  In case {\sl C4m} rotating at four
times the solar rate, in which ME at times exceeds KE
(Fig. \ref{timetrace}), angular velocity contrasts are always weaker than
in case {\sl Em} that rotates at the solar rate
(Fig. \ref{diffrotn_maghydro}). Similar damping of the differential
rotation by the Lorentz forces is also realized to a lesser extent in 3-D
MHD simulations of the solar convection zone (Brun et al. 2004; Brun 2004).

\begin{figure}[hpt]
\center
\epsscale{0.64}
\plotone{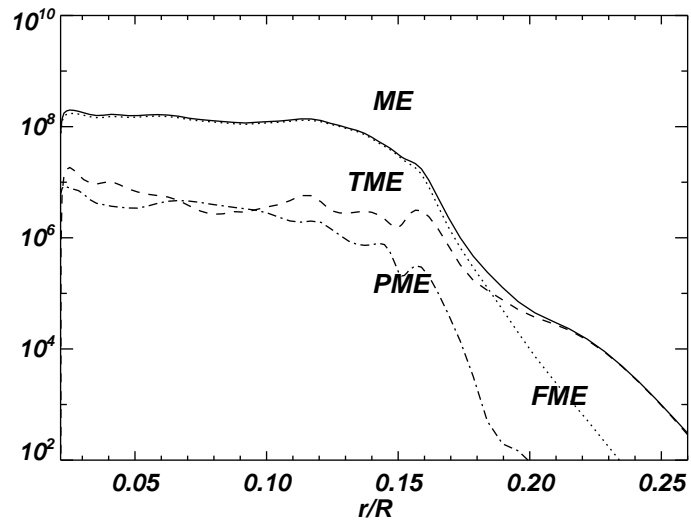}
\caption{\label{magbalance} Radial variation of  magnetic energy components
in case {\sl C4m}.
Shown are the energy in the mean
(axisymmetric) toroidal field (TME), the mean poloidal field (PME), and the
fluctuating (non-axisymmetric) fields (FME), together with their sum (ME), all averaged over radial
surfaces and in time.  In the convective core, FME  accounts for most of ME. Outside
the core, TME becomes the dominant component in the plummeting ME.}
\end{figure}

\begin{figure}[hpt]
\center
\epsscale{0.8}
\plotone{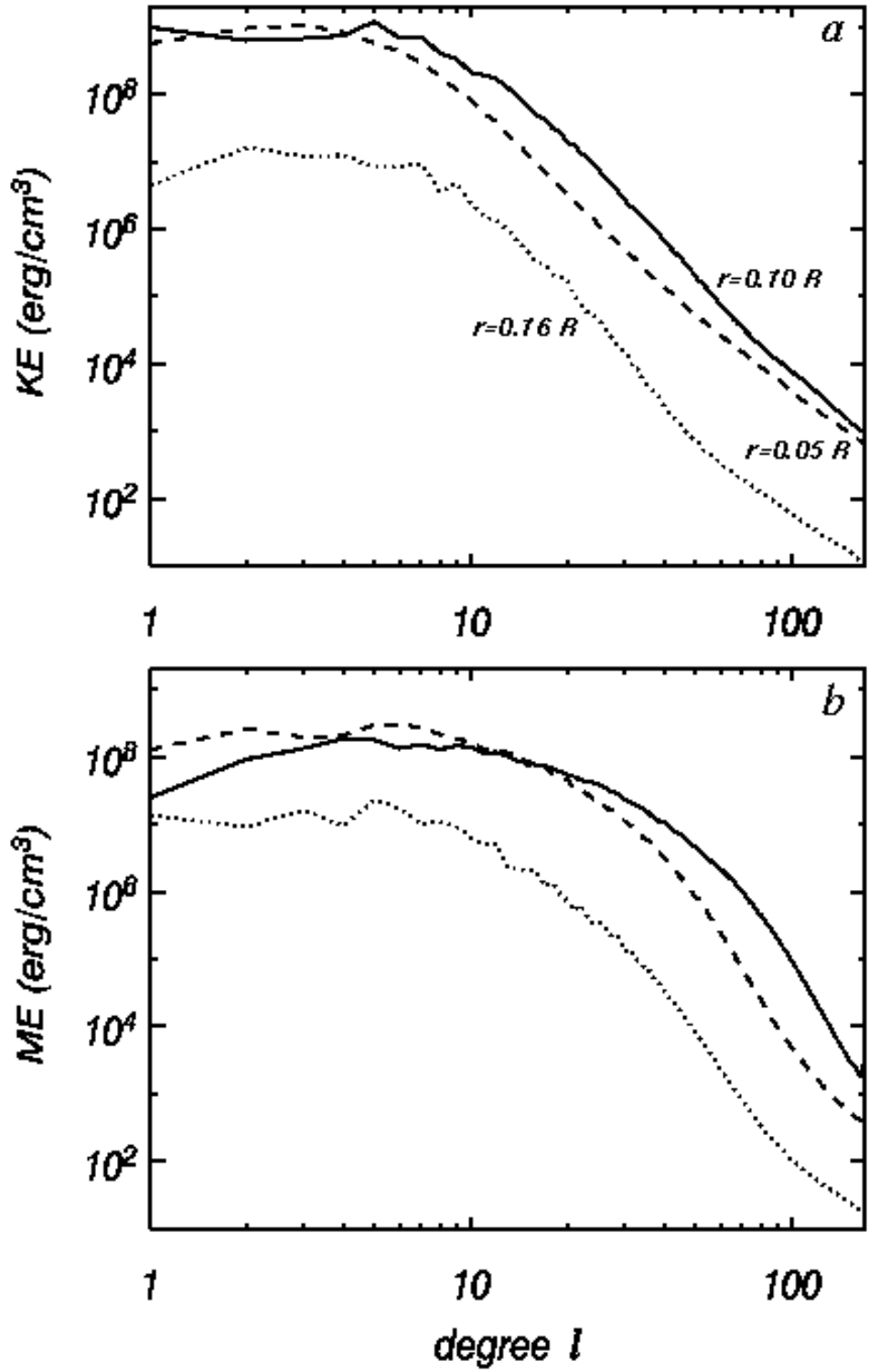}
\caption{\label{spectra} Time-averaged spectral distributions of ($a$)
kinetic energy (KE) and ($b$) magnetic energy (ME) with degree $\ell$ for case
{\sl Em}, evaluated on three spheres with radii indicated.}
\end{figure}

\begin{figure}[hpt]
\center
\epsscale{1.0}
\plotone{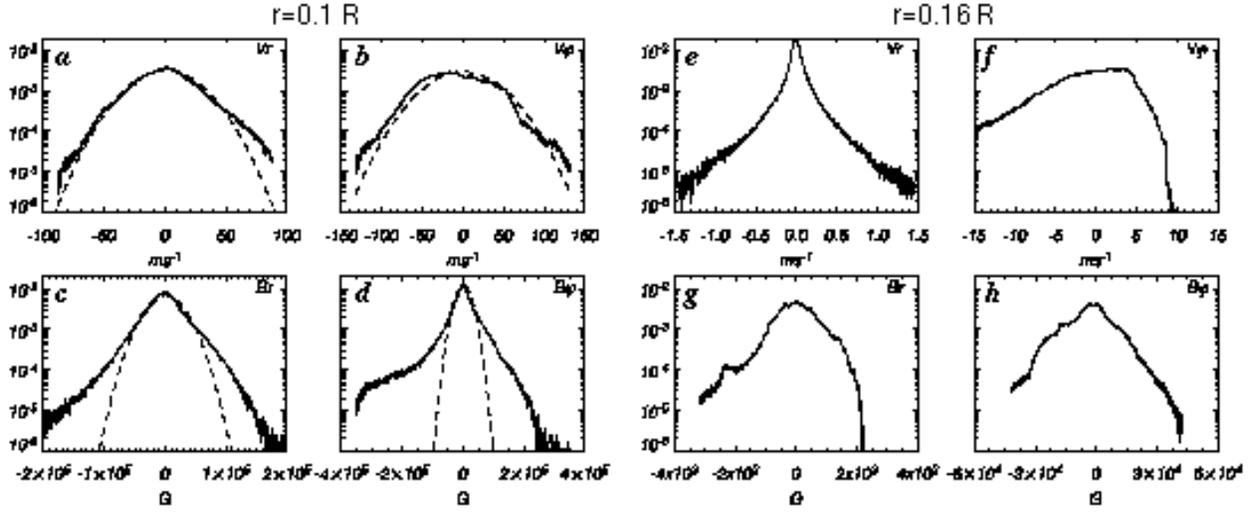}
\caption{\label{pdfs_Em} Time-averaged PDFs for case {\sl Em} of velocities
($v_r$ and $v_{\phi}$) and magnetic fields ($B_r$, $B_{\phi}$) sampled on
two spherical surfaces ($r=0.10 R$ and $r=0.16 R$).  Some Gaussian fits to
the distributions are indicated by dashed curves.}
\end{figure}

\begin{figure}[!ht]
\setlength{\unitlength}{1.0cm}
\begin{picture}(5,11.5)
\includegraphics{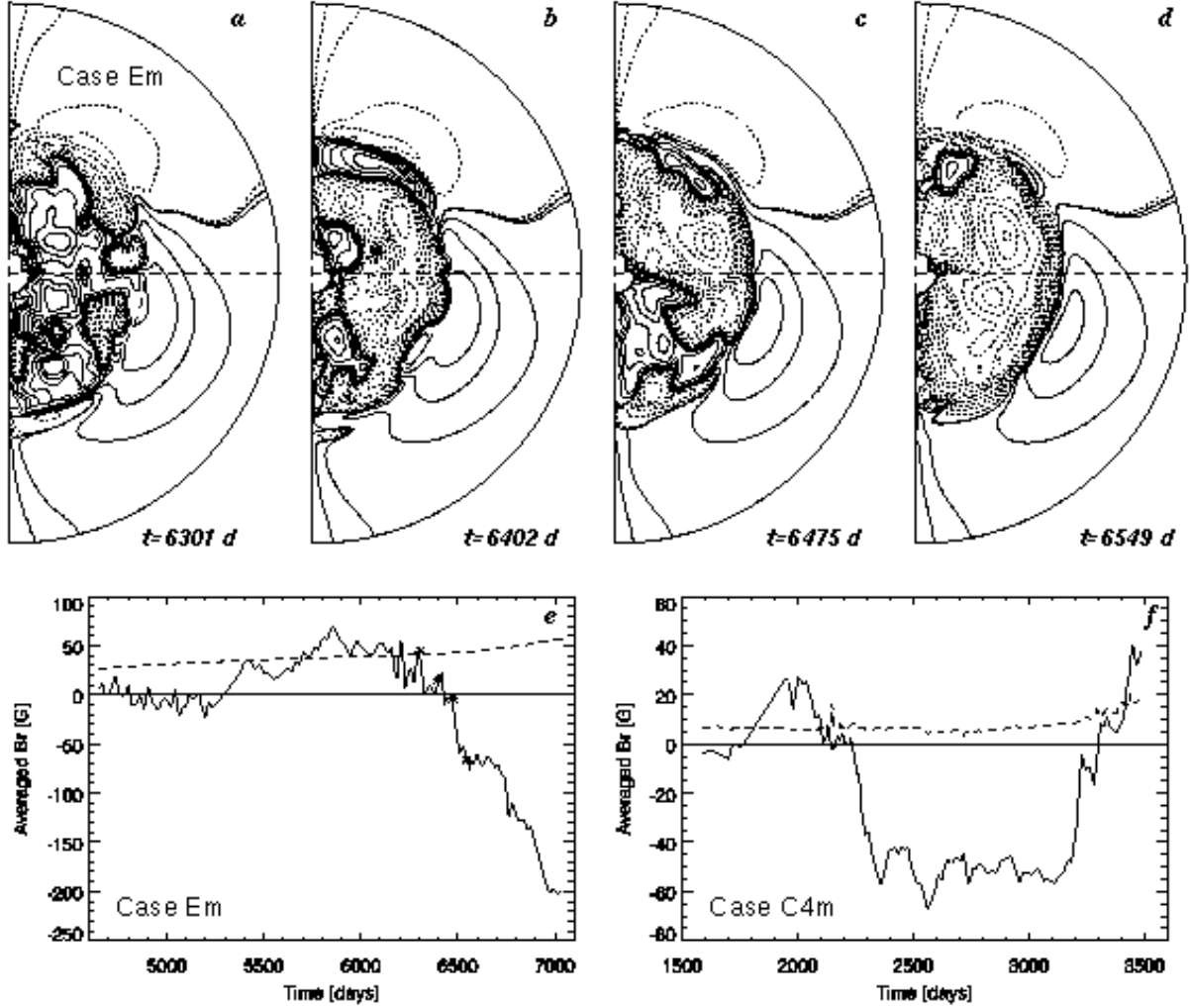}
\end{picture}
\caption[]{\label{Bpol} ($a$)--($d$) Temporal evolution of the axisymmetric
(mean) poloidal field $\left<{\bf B_p}\right>$ for case {\sl Em}, shown as meridional cross section at four selected times indicated.  Solid contours denote positive polarity
(field lines directed from north to south) and dotted contours denote
negative polarity.  ($e$) The accompanying radial field for case {\sl Em} at the convective core
boundary (solid line) and at the top of the domain (dashed) as averaged over
the northern hemisphere, shown over 2400 days late in the simulation with
timings of upper snapshots denoted. ($f$) Temporal evolution of average
polarity of $B_r$, as in ($e$), for case {\sl C4m}.}
\end{figure}

\begin{figure}[!ht]
\epsscale{1.0}
\plotone{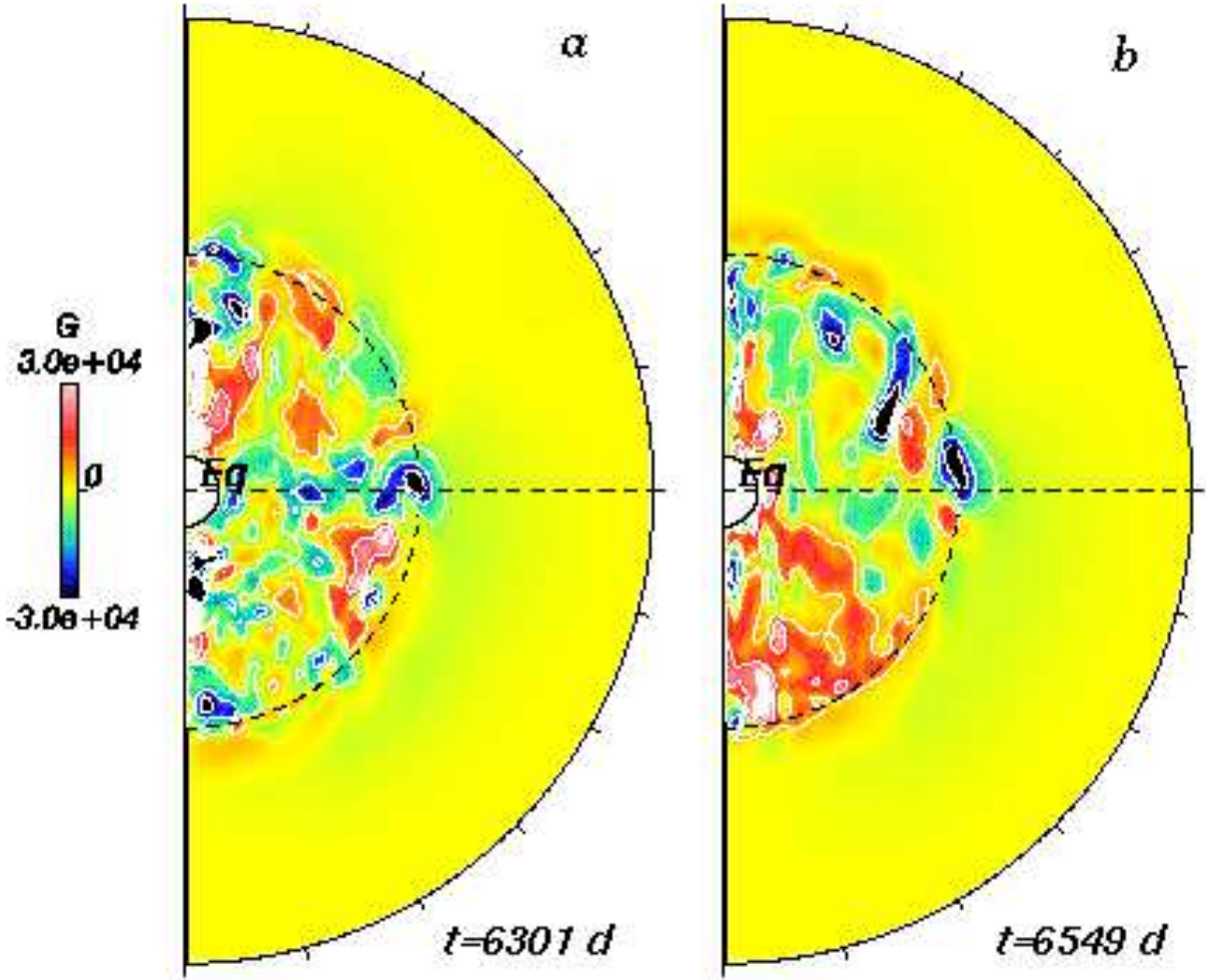}
\caption[]{\label{Btor} Variation of axisymmetric toroidal field $\left<B_t\right>$ 
with radius and latitude for case {\sl Em} at two instants
coinciding with Figs. \ref{Bpol}$a$,$d$.  Red and blue tones denote in turn
eastward (prograde) and westward (retrograde) field.}
\end{figure}

\begin{figure}[hpt]
\center
\epsscale{0.54}
\plotone{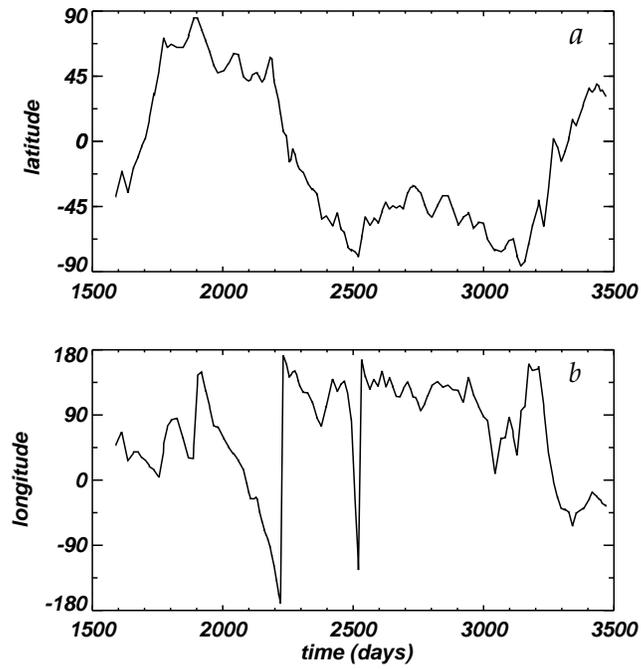}
\caption{\label{polewander} Wandering of the magnetic dipole axis in case
{\sl C4m} with time. ($a$) Its position with latitude as the field swings
between the northern and southern hemispheres.  ($b$) The gradual drift in
longitude of the positive pole.}
\end{figure}

How the magnetism is built and sustained by the flows is an intricate
matter.  Both the helical convection and the differential rotation have
significant roles, with axisymmetric fields generated in the core through
processes somewhat akin to the $\alpha$- and $\omega$-effects of mean-field
dynamo theory.  It would appear that the two effects contribute in roughly
equal measure to building axisymmetric toroidal fields within the core, as
indicated by MTE sampled at mid-core exceeding MPE by a factor of about two
in case {\sl Em} (Table 3).  The role of helical convection here in
generating field is unlike that prescribed in the simpler variants of
mean-field theory (i.e., the first-order smoothing approximation).  In
particular, the fluctuating field $B'$ in our simulations is not
proportional to the longitudinally-averaged field $<B>$, possibly because
the G-current is not small.  This may explain the absence of a linear
relationship between the fluctuating and axisymmetric fields in our
modeling.  In the radiative envelope, convective motions do not persist and
so cannot serve to build magnetism, but weak angular velocity contrasts
continue to generate axisymmetric toroidal magnetic fields through
stretching via the $\omega$-effect.  Thus MTE with increasing radius comes
to dominate over both FME and MPE, as seen in Figure \ref{magbalance}.

We also find that the magnetic energy peaks slightly at the bottom of the
convective domains due to the downward transport of magnetic fields by the
convection.  This is in sharp contrast with the much more peaked profile
found in our simulation of the solar dynamo (Brun et al. 2004), where the
asymmetry between weak upflows and strong downflows is more pronounced due
to a stronger density contrast there.

The intense and rapidly evolving magnetism realized within the core is
screened by the extensive radiative envelope, so assessing its possible
impact at the stellar surface is difficult.  The complex morphologies of
the magnetic fields, their periodic reversals in mean polarity, and their
intimate feedback upon the turbulent convection may all be hidden from
view.  If diffusion alone served to bring the magnetic fields outward from
the core, the rapid temporal variations (and so too the intricate spatial
structure) of the core fields would be obliterated by the characteristic
diffusive time scales of order millions of years.  

Whether the fields could migrate to the surface by means other than
diffusion, and thus perhaps contribute to the observed magnetism of Ap
stars, has been the subject of some debate.  Magnetic buoyancy
instabilities at the edge of the core could conceivably bring the fields to
the surface much more rapidly than diffusion.  We cannot address the rise
of such buoyant flux tubes directly in our simulations, since we model only
the interior portions of these stars at resolutions insufficient to capture
the highly concentrated structures needed for these instabilities to act.
However, MacGregor \& Cassinelli (2003) have used simple models to consider
how buoyant magnetic structures may traverse the radiative exterior.  They
deduce that magnetism from the core could arrive at the stellar surface in
less than the main-sequence lifetime of an A-type stars if the interior
fields were both very strong and highly fibril.  Further modeling by
MacDonald \& Mullan (2004) suggests that the presence of compositional
gradients would slow this process considerably.  Furthermore, the field
strengths likely to be realized at the surface from such buoyant flux tubes
are only modest (MacGregor \& Cassinelli 2003), in contrast with the kG
fields that are observed in some Ap stars.  The implications of these
recent studies remain somewhat unclear, for several effects that might
modify the rise of buoyant core magnetic fields have yet to be included,
among them global-scale circulations within the radiative envelope and the
twist and writhe of the flux tubes that could modify their stability.  The
strong surface fields also appear to occupy large sectors, which might be
very difficult to populate through the rise of individual elements.
Invoking fossil origins for the observed surface magnetism, with
predominate dipole structure surviving, is a favored explanation (e.g.,
Moss 2001) since strong fields could more easily be obtained.  The
interaction of the interior magnetism with such possible large-scale fossil
fields has also not yet been seriously studied. However, the alternative
possibility that the radiative envelope could induce a magnetic field via
dynamo action (Spruit 2002; MacDonald \& Mullan 2004) may encourage reconsideration of
that scenario.  Also, the generation by the overshooting convection of
internal waves could potentially play a role in the radiative zone,
creating shear layers that could subsequently amplify a magnetic field
(Kumar, Talon \& Zahn 1999).

Future work will thus be required to explore in detail the possible role of
the core convection dynamo in giving rise to the surface magnetism.
Likewise we have only briefly touched the possible
variations with rotation of the dynamo action and differential rotation.
Our limited sampling of two rotation rates provides some hints at that
variation (i.e., the faster case {\sl C4m} possesses a stronger magnetic
field than case {\sl Em}), but does not elucidate it, for we have
explored only one avenue in the vast parameter space that could be relevant
to real stars.  Indeed, the detailed nature of the flows and magnetism in
our simulations is surely affected by the many approximations we have made
in considering these stars.  How the far more turbulent flows attained in
actual stars impact the generation of magnetic fields and differential
rotation is quite uncertain. Yet some of the dominant features found here
may well turn out to be robust.  The conclusion that core convection drives
some form of sustained dynamo action, producing very strong fields that
feed back on the flows, appears to us to be inescapable.

This work was partly supported by NASA through SEC Theory Program grant
NAG5-11879, and through the Graduate Student Researchers Program
(NGT5-50416).  The simulations were carried out with NSF PACI support of
the San Diego Supercomputing Center (SDSC), the National Center for
Supercomputing Applications (NCSA), and the Pittsburgh Supercomputing
Center (PSC) as well as with the Centre de Calcul pour la Recherche et 
la Technologie (CCRT) of CEA at Bruy\`ere-le-Chatel and within the CNRS
supercomputers center IDRIS.  Much of the analyses 
were conducted in the Laboratory for Computational Dynamics (LCD) within JILA.
One of us (ASB) is grateful to R. Grappin and J.-P. Zahn for interesting discussions
and insights on turbulent MHD.

\begin{deluxetable}{ccc}
\tablecolumns{3}
\tablenum{1}
\tablecaption{Parameters for Magnetic Simulations}
\tablehead{
\colhead{Case} & \colhead{{\sl Em}} & \colhead{{\sl C4m}}}
\startdata
  &  Input parameters & \\
  \hline
  $N_r, N_{\theta}, N_{\phi}$ & 82, 256, 512 & 82, 256, 512 \\
  $\Omega_0$ (s$^{-1}$) & $2.6 \times 10^{-6}$ & $1.04 \times 10^{-5}$ \\
  $R_a$ & 3.1 $\times 10^5$ & 1.3 $\times 10^7$ \\
  $P_m$ & 5 & 5 \\
  $R_c$ & 0.33 & 0.12 \\
  $\nu$ (cm$^2$ s$^{-1}$) & $4.4 \times 10^{11}$ & $2.5 \times 10^{11}$  \\
  $\kappa$ (cm$^2$ s$^{-1}$) & $1.7 \times 10^{12}$ & $9.9 \times 10^{11}$ \\
  $\eta$ (cm$^2$ s$^{-1}$) & $8.7 \times 10^{10}$ & $5.0 \times 10^{10}$ \\
  $\tau_{\eta}$ (days) & 3900 & 6800 \\
  \hline
  & Measured at mid-depth of convective core& \\
  \hline
  $R_e$  & 160 & 210 \\
  $R_m$  & 800  & 1050  \\
  $\Lambda$ & 23.2 & 17.8 \\
  $P_e$ & 40 & 52 \\
  $R_o$  & $3.5 \times 10^{-2}$ & $6.2 \times 10^{-3}$ \\
\enddata
\tablecomments{The number of radial, latitudinal and longitudinal mesh points are $N_{r}, N_{\theta}, N_{\phi}$.
All simulations have an inner radius $r_{bot}=3.0 \times 10^{9}$ cm
and an outer radius $r_{top}=4.0 \times 10^{10}$ cm, with $L=1.7 \times
10^{10}$ cm the approximate radial extent of the convective core.  The
overall radius $R$ of the A-type star is $1.4 \times 10^{11}$ cm. The effective viscosity $\nu$, thermal
diffusivity $\kappa$, and magnetic diffusivity $\eta$ are quoted at the
middle of the convective core ($r=0.10R$), and likewise we evaluate there the Rayleigh number $R_a=(-\p
\bar{\rho}/\p S)\Delta S g L^3/\rho \nu \kappa$ (with $\Delta S$ the entropy
contrast across the core), the magnetic Prandtl number 
$P_m=\nu/\eta$, the convective Rossby number $R_c=\sqrt{R_a/T_a P_r}$, the Reynolds number
$R_e=\vvr' L/\nu$, the magnetic Reynolds number $R_m=\vvr' L/\eta$, the 
Elsasser number $\Lambda=\tilde{B}^2/4\pi\rho\eta\Omega_0$, the P\'eclet number
$P_e=R_eP_r=\vvr' L/\kappa$, the Rossby number 
$R_o=\tilde{v}^\prime/2\Omega_0 L$,
and the ohmic diffusion time $\tau_{\eta}=L^2/(\pi^2 \eta)$,
where $\vvr'$ is the rms fluctuating convective velocity and $\tilde{B}$ is the rms
magnetic field (see Table 2 and \S 2.3). An $R_e$ based on the peak velocity at mid-depth 
would be about a factor of 5 larger.  The Prandtl number 
$P_r = \nu/\kappa$ = 0.25 over the full depth range.}
\end{deluxetable}

\begin{deluxetable}{lllll}
\tablecolumns{5}
\tablenum{2}
\tablecaption{Velocity and Magnetic Field Amplitudes}
\tablehead{
\colhead{Case} & \colhead{Em} & \colhead{C4m} & \colhead{E} & \colhead{C4}
}
\startdata
 $\vrr$  & 20 & 15 & 26 & 19\\
 $\vtr$  & 22 & 16 & 23 & 16\\
 $\vphr$ & 22 & 21 & 43 & 72\\
 $\vphr'$ & 20 & 15 & 21 & 21\\
 $\vvr$ & 37 & 30 & 55 & 76\\
 $\vvr'$ & 36 & 26 & 38 & 32\\
 $\brr$ & 28 & 33 & - & - \\
 $\btr$ & 30 & 36 & - & -\\
 $\bphr$ & 28 & 45 & - & -\\
 $\bphr'$ & 27 & 44 & - & -\\
 $\bbr$ & 50 & 67 & - & -\\
 $\bbr'$ & 49 & 65 & - & -\\
\enddata
\tablecomments{Listed for both MHD simulations (cases {\sl Em}, {\sl C4m}) and their hydrodynamic
progenitors (cases {\sl E}, {\sl C4}) are the rms amplitude of the velocity
$\vvr$ and each of its components, $\vrr$, $\vtr$, and $\vphr$,
averaged over time and over a spherical surface at mid-depth in the convective 
core (at $r=0.10R$). Also listed are the rms amplitudes of the fluctuating
velocity $\vvr'$ and
its zonal component $\vphr'$, averaged in time and obtained after subtracting 
the longitudinal average.  We also indicate
(where appropriate) the corresponding rms amplitudes of the magnetic field and its
components, $\bbr$, $\brr$, $\btr$, $\bphr$, $\bbr'$, and $\bphr'$.
Velocities are expressed in m s$^{-1}$ and magnetic fields in $kG$.}
\end{deluxetable}

\begin{deluxetable}{lllll}
\tablecolumns{5}
\tablenum{3}
\tablecaption{Energy densities}
\tablehead{
\colhead{Case} & \colhead{Em} & \colhead{C4m} & \colhead{E} & \colhead{C4}
}
\startdata
 KE & 4.54$\times 10^7$ & 1.76$\times 10^7$  & 7.58$\times10^7$ &
1.67$\times 10^8$ \\
 DRKE/KE & 18.2\% & 28.3\% & 56.1\%  & 88.3\%\\
 MCKE/KE & 1.9\% & 0.2\% & 1.8\% & 0.2\% \\
 CKE/KE & 79.9\% & 71.5\% &  42.2\% & 11.5\% \\
 ME/KE & 28.3\% & 88.2\% & - & - \\
 MTE/ME\emph{c} & 3.4\% & 2.7\% & - & -  \\
 MPE/ME\emph{c} & 1.4\% & 2.5\% & - & -\\
 FME/ME\emph{c} & 94.4\% & 94.7\% & - & -\\
 ME\emph{c} & 1.27$\times 10^8$  & 2.20$\times 10^8$\\
 MTE/ME\emph{r} & 91.3\% & 98.1\% & - & -\\
 MPE/ME\emph{r} & 0.2\% & 0.1\% & - & -\\
 FME/ME\emph{r} & 9.6\% & 1.8\% & - & -\\
 ME\emph{r} & 6.85$\times 10^3$  & 4.93$\times 10^3$ & - & - \\
\enddata
\tablecomments{The kinetic energy density KE ($1/2 ~ \rb v^2$), averaged over volume
and time, is listed along with the relative contributions from
the convection (CKE), the differential rotation (DRKE), and the meridional circulation (MCKE),
together with the average magnetic energy density ME
($B^2/8\pi$) (where appropriate).  The relative contributions from each of the
components of ME, including the fluctuating field FME and
the axisymmetric $m=0$ toroidal and poloidal fields (MTE and MPE), are
evaluated both within the radiative zone (at $r=0.24R$, denoted
by $r$) and the convective core (at $r=0.10R$, denoted $c$), along with the
values of ME at those two depths.}
\end{deluxetable}

\clearpage

\clearpage

\pagebreak

\begin{figure}[hpt]
\center
\epsscale{0.65}
\plotone{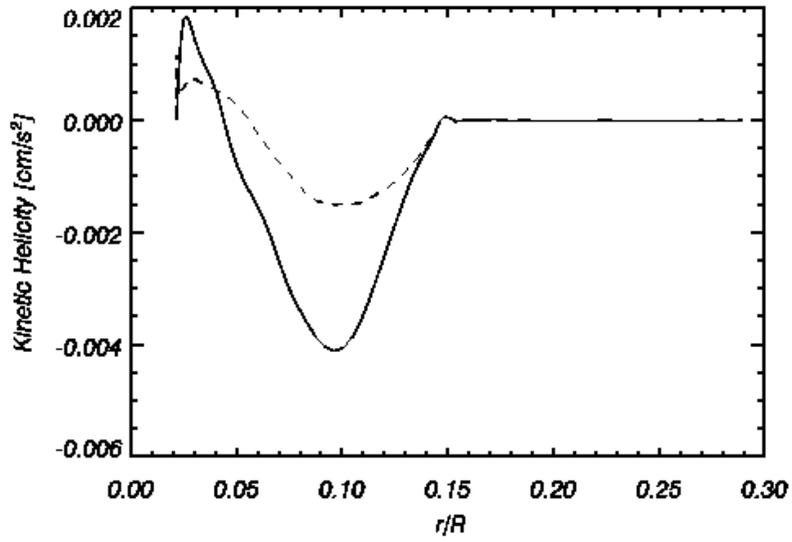}
\caption{\label{kinhel} Kinetic helicity as a function of radius for case {\sl Em} (dashed line)
and its purely hydrodynamical progenitor (solid line), as averaged over the
northern hemisphere and in time. 
The kinetic helicity is reduced in the presence of magnetism.
The kinetic helicity has been averaged over 120 days in both cases given
the large fluctuations that this
quantity undergoes in the convective core.}
\end{figure}

\end{document}